\documentclass[useAMS,usenatbib]{mn2e}
\usepackage[T1]{fontenc}
 \usepackage{pslatex}
\usepackage{amsmath}
\usepackage{amsfonts}
\usepackage{graphicx}

\usepackage{color}
\usepackage[normalem]{ulem}
\usepackage{fixltx2e}
\def\updated#1#2{{#2}}

{\newif\ifnotend
\notendtrue
\def\veclist{ABCDEFGHIJKLMNOPQRSTUVWXYZabcdefghijklmnopqrstuvwxyz.}
\def\top#1#2.{#1}
\def\tail#1#2.{#2.}
\loop\expandafter\xdef\csname v\expandafter\top\veclist\endcsname%
{{\noexpand\bf\expandafter\top\veclist}}
\edef\veclist{\expandafter\tail\veclist}
\if\veclist.\notendfalse\fi\ifnotend\repeat}

\def\d{{\rm d}}
\def\e{{\rm e}}
\def\aap{A\&A}
\def\apj{ApJ}
\def\apjs{ApJS}

\def\aj{AJ}
\def\mnras{MNRAS}
\def \physrep{Phys. Rep.}
\def\kms{{\,\rm km\,s^{-1}}}

\title[Nucleus of M31]{Three-dimensional Keplerian orbit-superposition models of the nucleus of M31}
\author[C. K. Brown and J. Magorrian]{C. K. Brown and  J. Magorrian\thanks{Email: \tt ckb@thphys.ox.ac.uk, magog@thphys.ox.ac.uk}\\
Rudolf Peierls Centre for Theoretical Astrophysics, 1 Keble Road, Oxford OX1 3NP}
\begin{document}

\pagerange{\pageref{firstpage}--\pageref{lastpage}} \pubyear{2013}

\maketitle

\label{firstpage}

\begin{abstract}
\noindent
We present three-dimensional eccentric disc models of the nucleus of
M31, modelling the disc as a linear combination of thick rings of
massless stars orbiting in the potential of a central black hole.
Our models are nonparametric generalisations of the parametric models
of Peiris \& Tremaine.
The models reproduce well the observed WFPC2 photometry, the detailed
line-of-sight velocity distributions from STIS observations along P1
and P2, together with the qualitative features of the OASIS kinematic
maps.

We confirm Peiris \& Tremaine's finding that nuclear discs aligned
with the larger disc of M31 are strongly ruled out.  Our optimal model
is inclined at $57^\circ$ with respect to the line of sight of M31 and
has position angle P.A.$= \theta_l + 90^\circ = 55^\circ$.  It has a
central black hole of mass $M_\bullet\simeq1.0\times10^8\,M_\odot$,
and, when viewed in three dimensions, shows a clear enhancement in the
density of stars around the black hole.  The distribution of orbit
eccentricities in our models is similar to Peiris \& Tremaine's model,
but we find significantly different inclination distributions, which
might provide valuable clues to the origin of the disc.
\end{abstract}

\begin{keywords}
galaxies: individual: M31 -- galaxies: nuclei -- galaxies: kinematics and dynamics
\end{keywords}

\section{Introduction}
\label{m31}

As the nearest spiral galaxy to our own, M31 allows the study of a
galactic centre in unmatched detail.  A particularly striking feature
of high-resolution $V$- and $I$-band photometry of the central few
arcseconds of M31 is its double nucleus  \citep{light74, lauer93, bacon94, king95}:
there are two peaks in surface brightness, the brighter of
which (known as P1) is extended and offset $\sim0\farcs5$ from the
fainter peak (known as P2), which is elongated and close to the
photometric centre of the bulge.
Early spectroscopic observations resolved steep gradients in the stellar
rotation velocities and a prominent peak in the stellar velocity
dispersion, which hints at the presence of a massive black hole
\citep{dressler88, kormendy88, bacon94, vdm94}.

\cite{tremaine95} put forward an elegant explanation for these
observations: the nucleus is a massive disc of red stars on eccentric,
nearly Keplerian, approximately aligned orbits around a central black
hole located at P2;  P1 is generated by orbital crowding of stars
lingering at apocentre.
Subsequent observations have been consistent with his model.
\cite{lauer98} observed the nucleus with the Wide Field Planetary
Camera 2 (WFPC2) on the corrected Hubble Space Telescope (HST),
confirming the bimodal structure.  \cite{kormendy99} obtained
spectroscopy with the Subarcsecond Imaging Spectrograph (SIS) finding
an asymmetric rotation curve and a constant colour across the nucleus,
showing P1 and P2's colours to be consistent with each other but not
the bulge or a globular cluster. Further kinematics were recorded with
the Faint Object Camera \citep{statleretal99}, the integral field spectrograph OASIS
\citep{bacon01} and the Space Telescope Spectroscopy Imaging
Spectrograph (STIS) on HST \citep{bender05}.  All of these
observations show that the kinematic centre of the nucleus is very
close to P2, as predicted by T95's eccentric disc model. 

Further refinement of this picture has come from studying the nucleus
at ultraviolet wavelengths.  Almost all of the UV emission from the
nucleus comes from a tiny ($<0\farcs1$) source located at P2
\citep{king95,lauer98,brown98}, whose optical--UV colours and spectra
are consistent with a population of A stars \citep{lauer98,bender05},
distinct from the red K-type spectrum of the rest of the nucleus
\citep{bender05}.  \cite{bender05} find that this compact, young, blue
population has a maximum velocity dispersion of $ \sigma =1183 \pm
201$ km s$^{ -1}$, significantly higher than the red stars' $ \approx
250$ km s$^{-1}$.  They label the UV peak P3 and have found that its
photometry and kinematics are well modelled by a separate,
almost-circular disc of blue stars around a central black hole of mass
($1.1$-$2.3)\times10^8 M_\odot$.  This provides very strong evidence
in support of the presence of a supermassive black hole, which is the
most fundamental requirement of T95's model.    Most recently,
\cite{lauer12} has found that the surface brightness profile of the young population is
described by an exponential profile of scale length $0\farcs075\pm0\farcs01$.

Tremaine's original eccentric disc model consisted of three Keplerian
orbits, coplanar with the main disc of M31, projected onto the sky and
convolved with Gaussian point spread functions (PSFs). The original
model included neither disc self-gravity nor a realistic treatment of
the disc's internal velocity dispersions but was still able to broadly
reproduce kinematic features. To be stable over many dynamical times
such a disc would require apsidal alignment to be maintained; T95
proposed that this could be achieved by the disc self-gravity and
argued that two-body relaxation would lead to a disc thickness of
$\approx 0.3$ times the disc radius. Estimates of the disc mass from
mass to light ratios place it at around $10^7 M_\odot$. This is
substantial enough to affect the dynamics of the disc although the
nucleus falls within the sphere of influence of the black hole, which
dominates the orbits and imposes regularity. A self-gravitating
eccentric disc mode maintaining orbital alignment would also precess
under a uniform pattern speed. Since the T95 model, several {\it
  two-dimensional} self-gravitating models with these properties have
been constructed \citep{statler99, bacon01, salow01, sambhus02,
  salow04}. These massive models have found a variety of disc masses,
pattern speeds and orbital distributions.
 
\cite{peiris03} took a different approach.  They revised the 1995
model and constructed fully three-dimensional models with a parametric
distribution function that ignored the self gravity of the disc; the
gravitational potential in their models is due solely to the central
black hole, which greatly simplifies the modelling procedure. Their
models are the most successful to date at fitting the observed
kinematics. They also found that thickened disc models that were
misaligned with respect to the large-scale M31 disc produced
significantly better fits than coplanar models, echoing a result seen
in the 2d models.

\begin{figure} 
\includegraphics*[width=84mm, trim = 100 0 100 0]{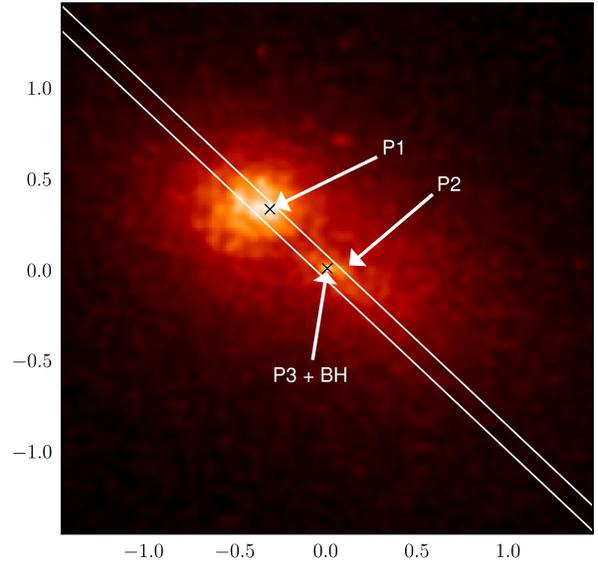}
\caption{$V$-band image of the nucleus
  \citep{lauer98} annotated to show the relative positions of P1, P2 and
  P3 in arcseconds. The pair of white lines represents the positioning and width of
  the STIS slit \citep{bender05}. The brightest point in the image is at P1 and is
  marked with cross. Centred in the image and marked with a cross is
  the location of the black hole and P3. P2 is on the anti-P1 side of
  the black hole. } \label{sample-figure}
\end{figure}

The current picture of the nucleus (see figure~\ref{sample-figure})
has the black hole \updated{}{(hereafter BH)} embedded in P3, which is
explained by a flat, circular exponential disc of blue stars
0\farcs033 from the photometric bulge centre.  This young, blue disc
is surrounded by the larger, red, eccentric disc.  P1 is made of stars
crowding at apoapsis, while P2 is now identified as stars at
pericenter in the elongated region on the anti-P1 side of P3. The
system remains of great interest: the origin of the nuclei and the
relationship between the red and blue populations are unexplained and
the mass and pattern speed of precession of the disc have not been
pinned down consistently. Understanding the dynamics of the disc will
help determine the BH mass more accurately which is of use in better
determining the relationship between a BH and the host galaxy. It is
also of interest to confirm whether the eccentric disc is aligned with
the main disc of M31.

In this paper we present a natural development of the modelling approach
started by \cite{peiris03}.  Like them, we ignore the self-gravity of
the disc, leaving the construction of fully self-gravitating models
for a subsequent paper.  \updated{Instead}{But instead} of considering parameterised
functional forms for the phase-space distribution function (hereafter
DF) of the nucleus, we model the DF non-parametrically as a mixture of
Gaussian rings whose amplitudes are allowed to vary: one of our goals is to
``let the data speak for themselves'' and then examine closely the structure of
 the DF.  The paper is organised as follows.  In section 2 we summarise the data we use and additional
post-processing applied.  Section 3 describes our modelling procedure
and section 4 our results.  Section 5 sums up.

Throughout this paper we adopt a distance of 770kpc to M31.

\section{Observational Data}

Our models use observations from three instruments: the HST photometry
recorded with WFPC2, detailed in Lauer et al 1998 (hereafter L98), the
high resolution kinematics from STIS (Bender et al 2005, B05) and the
kinematic maps from the integral field spectrograph OASIS (Bacon et al
2001, hereafter B01).

\subsection{Photometry}

We use the deconvolved $V$-band (F555W) WFPC2 photometry of L98.
This image is 1024$\times$1024 pixels in size with a pixel scale of
0\farcs011375. P3 is located at
(513, 518). We note that the orientation of the image has North
82.3$^\circ$ clockwise from the y-axis - the data has been rotated
away from the original alignment of the CCD, so that the corners of
the image are blank - but we make no attempt to reorient the
image. This image has already been reduced and deconvolved but we
perform some processing on the image in order to reduce this data to a tractable number
of observables for our fitting procedure. Bright foreground stars are
first masked out and we crop the image to a 512$\times$512 pixel region
centred on P3. This is a region of side 5\farcs824 and incorporates
the nucleus and the inner parts of the bulge and removes the blank
corners of the image. This cropped image is then rebinned to "super-pixels" of side $2^l$ pixels containing the mean of the $2^{2l} - n$ pixels, with $n$ the number of masked pixels falling in the super-pixel. We consider two schemes for choosing super-pixels: in the first we simply take $l=2$ everywhere so that super-pixels contain a $4 \times 4$ pixel region. In the other $l$ is chosen based on the surface brightness of the image scaled by $r^{-1/2}$, where $r$ is the radius from P3, to provide a good balance of detail around P2 and P1. This gives $l = 0$ (the resolution of the deconvolved image) at P1 and P2 and $l = 4$ in the outermost parts of the image. A total of 4096 super-pixels are generated in this scheme.

Ajhar et al (1997) and Lauer et al (1998) showed that that at the
resolution of WFPC2 the underlying number of stars in M31 shows strong
surface brightness fluctuations. The limited number of stars is the
dominant residual between the structure of the nucleus and any model
we fit. We treat the noise from the fluctuations as the sole source of our
errors in our photometry and ignore photon noise so that the
fractional error in a spatial bin is given by $\sigma =
\bar{N}^{-1/2}$, where $\bar{N}$ is the effective number of stars
within the spatial bin. L98 gives the effective magnitude of each
SBF ``star'' as $\bar{m}_I = 23.4$, or $224\,L_\odot$.  The error associated
with a spatial bin containing $n\,L_\odot$ is therefore $\sqrt{224n}\,L_\odot$.

\subsection{STIS kinematics}

The STIS kinematics are taken from Table 5 of \cite{bender05}. These consist of
the bulge-subtracted Gauss--Hermite coefficients derived with the
Fourier Correlation Quotient Method at the 22 positions listed for
each of $V$, $\sigma$, $h_3$ and $h_4$ and their accompanying
errors. We adopt the same slit widths of 0.1\arcsec and position angle
of 39$^\circ$.  The quoted positions are given with respect
to the location of P3.

We use equation (21) of \cite{peiris03} to model the STIS PSF.  When
this is convolved with a a $0\farcs1$-square top hat to represent the slit,
the result is well approximated by a double Gaussian
\begin{equation}
\mathop{\hbox{PSF}}(x, y) = \frac{1}{\sum_i I_i}\sum_{i=1}^2\frac{I_i}{2\upi\sigma_i^2}e^{-\frac{(x^2 + y^2)}{2\sigma_i^2}}
\label{eq:STISPSF}
\end{equation}
with amplitudes $I_1=0.24$, $I_2=0.76$ and dispersions $\sigma_1=0\farcs042$
and $\sigma_2=0\farcs087$ respectively.  We assume that
equation~\eqref{eq:STISPSF} as the effective PSF of the STIS
observations.  A more sophisticated treatment might take account of
asymmetries in the PSF and the variations in spatial binning along the
slit.

\subsection{OASIS kinematics}

We also make use of the kinematics of the integral field spectrograph
OASIS, which were kindly provided by Eric Emsellem. We opt to use the higher S/N data set,
M8. This consists of $V$, $\sigma$, $h_3$ and $h_4$ values derived
from spectra taken at 1123 positions, spaced by 0\farcs09. We
registered the image with the WFPC2 data using a similar process to
B01.  We found a close match in angle
(0.7$^\circ$) and a small offset of (-0\farcs02, -0\farcs03) between
the two images.

Like B01 we assumed that the OASIS measurements have a PSF that
can be described by a sum of three Gaussians and allowed the
parameters of these Gaussians to float freely in the registration
process, taking the form
\begin{equation}
\mathop{\hbox{PSF}}(x, y) = \frac{1}{\sum_i I_i}\sum_{i=1}^3\frac{I_i}{2\upi\sigma_i^2}e^{-\frac{(x^2 + y^2)}{2\sigma_i^2}}
\label{eq:OASISPSF}
\end{equation}
The resulting PSF differs from that found by B01 with
$\sigma_1 = 0\farcs230$, $\sigma_2 = 0\farcs587$, $\sigma_3 = 0\farcs440$ and $I_2/I_1 =
0.836$ and $I_3/I_1 = 0.057$.

\section{Modelling procedure}

Our models are straightforward generalisations of those of PT.  In
particular, we ignore the self gravity of the disc and the
gravitational influence of the bulge and assume that the potential is
purely Keplerian.  The mass of the central black hole $M_\bullet$ is
the single free parameter in our model potential.  We assume that the
BH is located at P3. 

\subsection{Coordinate systems}

Following PT03 we use three different coordinate systems: ``orbit
plane'', ``disc plane'' and ``sky plane''.  All three coordinate
systems have origin $O$ coincident with the BH. Our models have
biaxial symmetry.  This provides a natural definition of the ``disc
plane'' $(x,y,z)$ coordinate system: the model is symmetric under
reflections in the $(x,y)$ and \updated{$(y,z)$}{$(x,z)$} planes.  The orbit-plane and
sky-plane coordinate systems are defined as follows. In the potential
of the BH all orbits are Keplerian ellipses.  An orbit with semi-major
axis $a$ and eccentricity $e$ defines a coordinate system $(x',y',z')$
in which the $Oz'$ axis is parallel to the orbit's angular momentum
vector and the $Ox'$ axis points towards pericentre.  That is,
\begin{equation}
\begin{split}
x' =&\;a(\cos E - e),\\
y' =&\;a\sqrt{1-e^2}\sin E,\\
z' =&\;0,
\end{split}
\label{eq:pos}
\end{equation}
where the eccentric anomaly $E$ is related to the mean anomaly $M$
through Kepler's equation $M=E-e\sin E$.  In this coordinate system
the apocentre of the orbit is located at $(x',y',z')=(-a(1+e),0,0)$
and the pericentre at $(a(1-e),0,0)$.  Each star defines its own
orbit-plane $(x',y',z')$ coordinate system.

A star's disc-plane coordinates $(x,y,z)$ are related to its orbit-plane coordinates $(x',y',z')$ through
\begin{equation}
\begin{split}
\begin{pmatrix}
x \\  y \\ z
\end{pmatrix}
&=
\begin{pmatrix}
\cos \Omega & -\sin \Omega & 0 \\
\sin \Omega & \cos \Omega & 0\\
0 & 0 & 1
\end{pmatrix}
\begin{pmatrix}
    1 & 0 & 0 \\
0 & \cos I & -\sin I \\
0 & \sin I & \cos I
\end{pmatrix}\\
&\qquad\times
\begin{pmatrix}
\cos \omega & -\sin \omega & 0 \\
\sin \omega & \cos \omega & 0\\
0 & 0 & 1
\end{pmatrix}
\begin{pmatrix}
x' \\
y' \\
z' \end{pmatrix},
\end{split}
\end{equation}
where the angles $\omega$, $I$ and $\Omega$ are the star's argument of
periapsis, the inclination and longitude of the ascending node,
respectively.
Any orbit in the Keplerian potential of the BH can be labelled by the
five integrals of motion $(a,e,\omega,I,\Omega)$, but it proves
convenient to replace $e$ and $\omega$ by the 
eccentricity vector $\ve\equiv
(e\cos\varpi, e\sin\varpi, 0) = (e_x, e_y, 0)$, where 
the longitude of periapsis $\varpi = \omega + \Omega$.
The vector $\ve$ points from the BH towards the projection of the
pericentre onto the $z=0$ disc plane.  Its magnitude is the scalar
eccentricity~$e$ of the orbit.

Projected, sky-plane coordinates $(X,Y,Z)$ are related to disc plane coordinates via 
\begin{equation}
\begin{split}
\begin{pmatrix}
X \\  Y \\ Z
\end{pmatrix}
&=
\begin{pmatrix}
\cos \theta_l & -\sin \theta_l & 0 \\
\sin \theta_l & \cos \theta_l & 0\\
0 & 0 & 1
\end{pmatrix}
\begin{pmatrix}
    1 & 0 & 0 \\
0 & \cos \theta_i & -\sin \theta_i \\
0 & \sin \theta_i & \cos \theta_i
\end{pmatrix}\\
&\qquad\times
\begin{pmatrix}
\cos \theta_a & -\sin \theta_a & 0 \\
\sin \theta_a & \cos \theta_a & 0\\
0 & 0 & 1
\end{pmatrix}
\begin{pmatrix}
x \\
y \\
z
\end{pmatrix},
\end{split}
\label{eq:eulermat}
\end{equation}
in which $(\theta_a,\theta_i,\theta_l)$ are the three Euler angles
specifying the orientation of the disc with respect to the observer's
reference frame.  The $(X,Y)$ plane is the sky plane, with the
positive $X$ axis pointing west and positive $Y$ axis north.  The $Z$
axis then points {\it towards} the observer; the line-of-sight
velocity is therefore $V_{\rm los}=-\dot Z$, following the usual
convention that that receding objects have $V_{\rm los}>0$.

\subsection{Distribution function}

We model only the old stellar population of the disc; the young stars
around P3 and the bulge are treated as contaminants (see
\S\ref{sec:bulge} below).  Assuming that the old stars are
collisionless and homogenous, their dynamics can be completely
described by a distribution function (DF) $f(\vx,\vv)$, defined such
that $f(\vx,\vv)\,\d^3\vx\,\d^3\vv$ is the expected number of such
stars within a small phase-space volume $\d^3\vx\,\d^3\vv$ around the
point $(\vx,\vv)$.
We further assume that the system is in a steady state.  By Jeans'
theorem the DF must be expressible as a function of the integrals of
motion $(a,\ve,I,\Omega)$.  The number of stars within a small
phase-space volume element $\d^3\vx\d^3\vv$ is then
\begin{equation}
\label{eq:dos}
f(\vx,\vv)\d^3\vx\d^3\vv = f(a,\ve,I)\,
\frac{1}{2}(GM_\bullet)^{3/2}a^{1/2}\sin I\d I\,\d a\d\ve\d\Omega\d M,
\end{equation}
the $\frac{1}{2}(GM_\bullet)^{3/2}a^{1/2}\sin I$ factor coming from the Jacobian relating
$(\vx,\vv)$ to $(a,\ve,I,\Omega,M)$.

We assume that the distribution function can be decomposed into a
weighted sum of rings,
\begin{equation}
f = \sum_j{w_j f_j(a,\ve, I)},
\label{eq:ringsum}
\end{equation}
in which each ring has a uniform distribution of $\Omega\in[0,2\upi)$
and Gaussian distributions in $a$,
$\ve$, $I$\updated{}{,}
\begin{equation}
\begin{split}
f_j(a, \ve, I) = &N_j\exp{\left[-\frac{(a-a_j)^2}{2{\sigma_{a,
       j}}^2}\right]}\exp{\left[-\frac{(\ve-\ve_j)^2}{2{\sigma_{\ve,
       j}}^2}\right]}\times\\
&\exp{\left[-\frac{I^2}{2{\sigma_{I, j}}^2}\right]},
\end{split}
\label{eq:ring}
\end{equation}
and it is understood that $f_j(a,\ve,I)=0$ if $a<0$ or $|\ve|>1$.  The
normalisation factor $N_j$ is included to give each ring unit total
luminosity.  We use a Monte Carlo method to construct the rings, which
avoids explicit calculation of $N_j$.

Our rings span a $(20\times9\times4)$ grid in $(a_j, e_{x,j},
\sigma_{I,j})$, with mean semimajor axis $a_j$ running logarithmically
from $0\farcs03$ to $10''$, mean $x$ component of eccentricity vector
$e_{x,j}$ from $-0.8$ to $+0.8$ in steps $\Delta e=0.2$, and the
dispersion in inclination $\sigma_{I,j}$ drawn from $\{12^\circ,
24^\circ, 36^\circ, 48^\circ\}$.  For the spreads in $(a,e)$ we take
$\sigma_{a,j}=0.8a_j \Delta \log a$ and $\sigma_{e,j}=0.8\Delta
e=0.16$.  All rings have mean $e_{y,j}=0$. Together with the zero mean
inclination of each ring this makes our models symmetric under
reflection in the $y=0$ and $z=0$ planes.  Rings with mean $e_x>0$ are
``aligned'' in the sense that their apocentre (and therefore peak
density) lies somewhere along the negative $x$ axis.  Rings with mean
$e_x<0$ have the opposite orientation.

\subsection{Observables of each ring}

\label{sec:ringobs}
We compute the projected properties of each ring $f_j$
(equ.~\ref{eq:ring}) by sampling it with $10^6$ points, drawing $N_o =
5000$ values of $(a,\ve,I, \Omega, \omega)$ from the distribution
$f_j(a,\ve,I)a^{1/2}\sin I$ that appears on the right-hand side
of~\eqref{eq:dos}, then sampling 200 points equispaced in mean
anomaly~$M$ along each of these orbits.  To avoid sampling artefacts
we select the initial value of $M$ for each orbit at random.

\subsubsection{WFPC}
We use this sample of $10^6$ points to estimate the projected surface density
(or, equivalently, the zeroth velocity moment)
\begin{equation}
\mu^{0,\rm fine}_j(X,Y)\equiv\int \d Z \d V_X\d V_Y\d V_Z\, f_j
\label{eq:mu0fine}
\end{equation}
on a fine grid of $0\farcs011375\times0\farcs011375$ pixels on the
$(X,Y)$ sky plane by simply counting the number of points that fall
within each pixel. The contribution of the $j^{\rm th}$ ring to each
WFPC2 ``superpixel'' is found by taking the mean value of $\mu^{0,\rm
  fine}_j(X,Y)$ over the relevant range of $(X,Y)$.  We do not carry
out any PSF convolution for the WFPC2 photometry because the image of
\cite{lauer98} has already been deconvolved.

\subsubsection{OASIS}

The OASIS kinematics of \cite{bacon01} consist of measurements of mean
velocity $V(X,Y)$ and velocity dispersion $\sigma(X,Y)$, together with
Gauss-Hermite coefficients $h_3(X,Y)$ and $h_4(X,Y)$.  We ignore their
$h_3$ and $h_4$ measurements and assume that their measured $V(X,Y)$
and $(V^2+\sigma^2)(X,Y)$ distributions probe directly the
(PSF-convolved) first- and second-order moments of the line-of-sight velocity distribution (LOSVD) at the
point $(X,Y)$ on the sky.  This is a good approximation provided the
underlying PSF-convolved LOSVDs are reasonably close to Gaussian.  For
modelling purposes it is more natural to consider luminosity-weighted
moments
\begin{equation}
  \begin{split}
 \mu^1(X,Y)&\equiv I(X,Y)V(X,Y),\\
 \mu^2(X,Y)&\equiv   I(X,Y)(V^2+\sigma^2)(X,Y),
  \end{split}
\end{equation}
where $I(X,Y)$ is the underlying surface
brightness.  We obtain $I(X,Y)$ by convolving the WFPC2 image with the
OASIS PSF~\eqref{eq:OASISPSF}.  The observational errors on these
first- and second-order luminosity-weighted velocity moments
are obtained by adding the uncertainties on $I$, $V$ and $\sigma$ in
quadrature in the obvious way:
\begin{equation}
  \begin{split}
 (\Delta\mu^1)^2&=(\Delta I)^2V^2+I^2(\Delta V)^2\\
(\Delta\mu^2)^2&=(\Delta I)^2(V^2+\sigma^2)+I^2(2|V|\Delta
V)^2+I^2(2\sigma\Delta\sigma)^2.   
  \end{split}
\end{equation}

Following~\eqref{eq:mu0fine} above, the contribution of the $j^{\rm
  th}$ ring to the luminosity-weighted
first and second moments of the line-of-sight velocity distribution,
\begin{equation}
\begin{split}
\mu^{1,\rm fine}_j(X,Y)&\equiv\int \d Z \d V_X\d V_Y\d V_Z \,V_Z f_j,\\
\mu^{2,\rm fine}_j(X,Y)&\equiv\int \d Z \d V_X\d V_Y\d V_Z \,V_Z^2 f_j,
\end{split}
\end{equation}
are estimated by weighting each of the $10^6$ sample points by $V_Z$
and $V_Z^2$, respectively.  Having these $\mu^{i,\rm fine}_j(X,Y)$
distributions we convolve with the OASIS PSF~\eqref{eq:OASISPSF} to
obtain the contribution the ring makes to the model's predictions for
the first and second velocity moments of the OASIS kinematics.

\subsubsection{STIS}

\label{sec:stismodmom}

\begin{figure}
  \includegraphics*[width=\hsize, trim = 40 20 40 20]{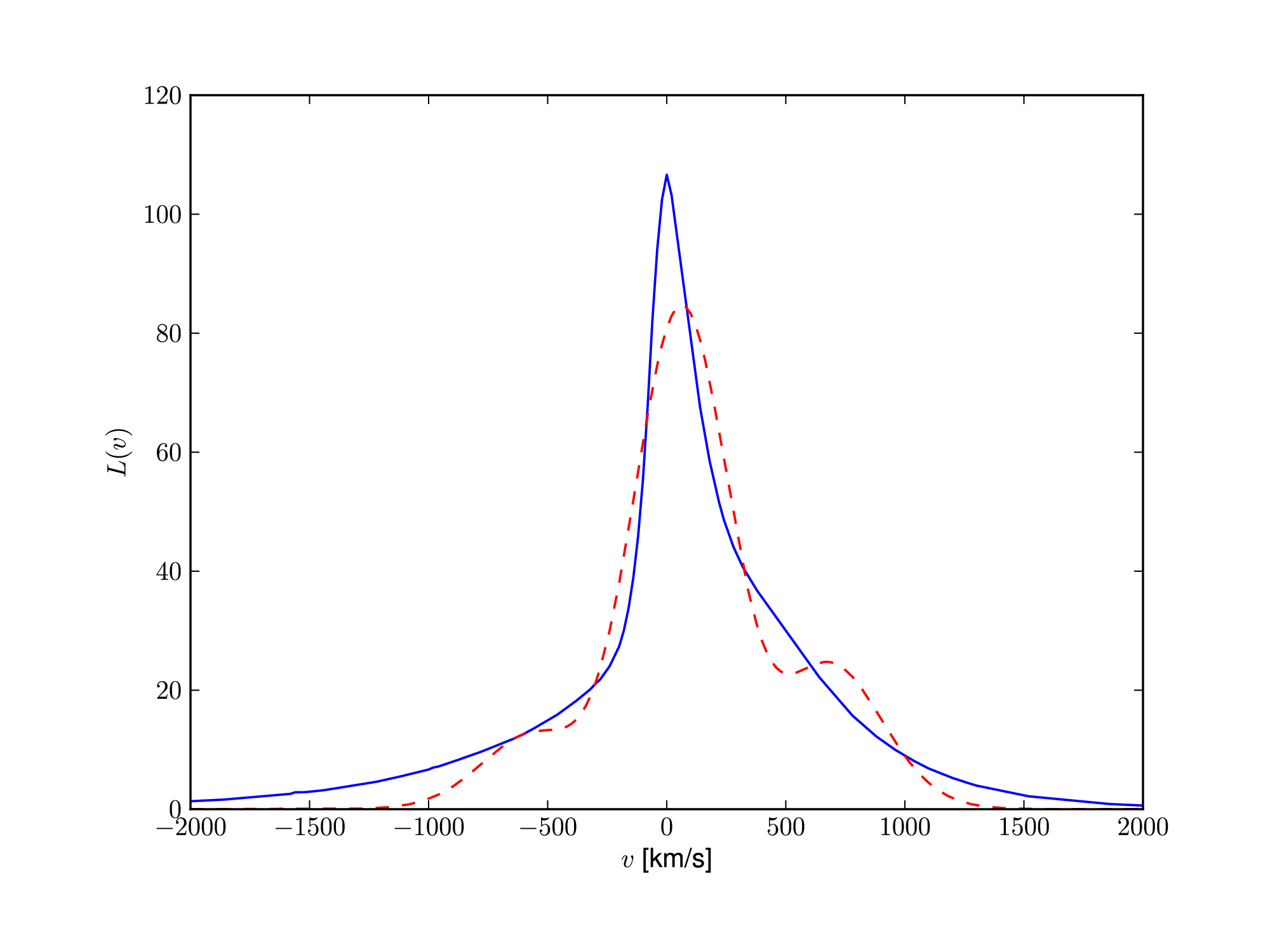}
  \caption{LOSVD of a typical model at offset $R=0\farcs2$ along the
    STIS slit (solid blue curve) and its 4-th order Gauss--Hermite
    reconstruction (equ~\ref{eq:GHint100}, dashed red curve).  The best-fit Gaussian to the
    LOSVD has $(\gamma,V,\sigma)=(0.81,-85\kms,281\kms)$.  For
    comparison, the classical zeroth-, first- and second-order  moments of this LOSVD are
    1, $V_0=-36\,\rm km\,s^{-1}$ and $\sigma_0=759\,\rm km\,s^{-1}$, the
    latter being dominated by the strong high-velocity wings.  }
  \label{fig:typlosvd}
\end{figure}

Our first attempt at fitting the STIS kinematics was based on the same
assumption that we could use the STIS $V$ and $\sigma$ measured by
B+05 as direct estimates of the (STIS PSF-convolved) first- and
second-order moments of the LOSVD.  That did not work well: at STIS
resolution the LOSVDs are far from Gaussian, as we show on
Figure~\ref{fig:typlosvd}, and the strong high-velocity wings
caused by the BH mean that it is not possible to obtain reliable
estimates of the first and second moments from the observed
spectra. This last point was one of the motivations for the
introduction of Gauss--Hermite series \citep{vdm93} to parametrize LOSVDs.
Therefore we fit our models to B+05's Gauss--Hermite parametrizations
of the LOSVDs along the STIS slit.  Our method for fitting the Gauss--Hermite coefficients
follows the same lines used in other orbit-superposition models (e.g.,
\cite{cretton99}), but taking extra care to treat the
normalisation of the LOSVDs correctly.

We recall some details of Gauss--Hermite expansions.  Suppose that we are
given an LOSVD $L_0(v)$, normalised such that $\int L_0(v)\,\d v=1$.
The Gauss--Hermite expansion of this $L_0(v)$ is
\begin{equation}
  L(v|\gamma,V,\sigma) = \frac{\gamma\alpha(w)}{\sigma}\sum_{j=0}^\infty h_jH_j(w),
  \label{eq:GHseries}
\end{equation}
where $w\equiv (v-V)/\sigma$, $\alpha(w)=\e^{-w^2/2}/\sqrt{2\upi}$ is
the standard Gaussian and the $H_j(w)$ are Hermite polynomials.  We
adopt vdMF93's normalisation for the latter.  Using the orthogonality
properties of the $H_j$, it is easy to show that the Gauss--Hermite
coefficients $h_l$ are given by
\begin{equation}
  h_l(\gamma,V,\sigma)=\frac{2\sqrt\upi}\gamma\int_{-\infty}^\infty L_0(v)H_l(w)\alpha(w)\d v.
\label{eq:GHint}
\end{equation}
That is, there is a different Gauss--Hermite series~\eqref{eq:GHseries} for each
choice of $(\gamma,V,\sigma)$.  The ``line-strength'' parameter
$\gamma$ simply scales all the $h_j$, but it proves important as we shall
now see.

A particularly natural choice of the parameters $(\gamma,V,\sigma)$ are
those that minimise 
\begin{equation}
  \chi_0^2 = \int_{-\infty}^\infty \left[L_0(v)-\frac{\gamma\alpha(w)}\sigma\right]^2\,\d v,
\label{eq:chisq0}
\end{equation}
in which case it can be shown that the first few Gauss--Hermite
coefficients from~\eqref{eq:GHint} become
$h_0=1$ and $h_1=h_2=0$.  In other words, if we choose
$(\gamma,V,\sigma)$ to be the parameters of the best-fit Gaussian to
the LOSVD~$L(v)$, then the LOSVD can be written as
\begin{equation}
  L(v)=\frac{\gamma\alpha(w)}\sigma\left[1+\sum_{j=3}^\infty h_jH_j(w)\right],
\label{eq:GHint100}
\end{equation}
with $h_3$, $h_4$, ... given by the integral~\eqref{eq:GHint}.
Conversely, if we adopt the parametrization~\eqref{eq:GHint100} and
fit $(\gamma,V,\sigma,h_3,h_4,...)$ simultaneously to $L_0(v)$, then
we get back the same parameters we would obtain by first fitting
$(\gamma,V,\sigma)$ by minimising~\eqref{eq:chisq0} and then
using~\eqref{eq:GHint} to find the~$h_i$.  For a strongly non-Gaussian
$L_0(v)$ the parameters $(\gamma,V,\sigma)$ obtained by
minimising~\eqref{eq:chisq0} need not be close to zeroth-, first- and
second-order moments of $L_0(v)$, as shown on
Figure~\ref{fig:typlosvd}.  In particular, $\gamma$ need not be close
to~1.

Gauss--Hermite fits to the LOSVDs of real galaxies, including B+05's
measurements of M31, adopt the parametrization~\eqref{eq:GHint100} and
fit $(\gamma,V,\sigma,h_3,h_4,...)$.  Unfortunately, the parameter
$\gamma$ is rarely reported, presumably because it is strongly
affected by systematic effects in the fitting procedure, such as
template mismatch, and because it does not affect the shape or width
of the LOSVDs.  Nevertheless, we note that it is an essential part of
any Gauss--Hermite expansion.  For now we assume that $\gamma$ is
known.  Our iterative scheme for reconstructing it from the models is
described in section~\ref{sec:gamma} below.

Notice that equation~\eqref{eq:GHint} shows that the
Gauss--Hermite coefficients $h_l$ can be thought of as {\it modified
  moments}: each $h_l$ is the integral over velocity space of some
linear combination of the classical moments $1,w,w^2,...,w^l$, but
weighted by the Gaussian factor~$\alpha(w)$.
Therefore, given a Gauss--Hermite fit $(\gamma,V,\sigma,h_{3},h_{4})$ to the
line-of-sight velocity distribution at projected position $(X,Y)$, we
treat $(\gamma,V,\sigma)$ as being known perfectly and take the
(luminosity-weighted) modified moments $(\tilde
\mu^0_{\rm obs},...,\tilde\mu^4_{\rm obs})\equiv I\times (1,0,0,h_{3},h_{4})$ as our
observables, where the surface brightness $I(X,Y)$ is obtained by
convolving the WFPC2 image by the STIS PSF~\eqref{eq:STISPSF}.  We  use
equations (10) of vdMF93 to propagate B+05's quoted uncertainties
$(\Delta V_i,\Delta\sigma_i,\Delta h_{3,i},\Delta h_{4,i})$ to our
observational errors $(\Delta\tilde\mu^0_{i,\rm obs},...,\Delta\tilde
\mu^4_{i,\rm obs})$.

Just as for the classical velocity moments, for each ring $j=1,2,...$ we
use our Monte Carlo sample of $10^6$ positions and velocities to
evaluate the expressions
\begin{equation}
\begin{split}
\tilde\mu_j^{k,\rm fine}(X,Y|{\gamma},V_i,\sigma_i)&\equiv
\frac{\sqrt2}{\gamma_i}\int \d Z \d V_X\d V_Y\d V_Z\\
&\quad\times \exp\left[-\frac12w^2\right]H_k(w)\, f_j
\end{split}
\label{eq:modmom}
\end{equation}
for the modified moments,
in which the rescaled velocity
\begin{equation}
w \equiv \frac{(-V_Z)-V_i}{\sigma_i}.
\end{equation}
Then we convolve each of these distributions with the STIS
PSF~\eqref{eq:STISPSF}  and
read off the values at $(X,Y)=(X_i,Y_i)$, giving the contribution
of the $j^{\rm th}$ ring to the $k^{\rm th}$
modified moment of the $i^{\rm th}$ STIS data point,
$\tilde\mu^k_{i,\rm obs}$.

\subsection{Modelling the effects of the bulge and P3}

\label{sec:bulge}

Our ring system is designed to model only the old red stars of the
eccentric disc, but some of the light observed in the central few
arcsec of M31 comes from other sources.  The two main contaminants are
M31's bulge and the compact young stellar cluster at P3.

We follow \cite{kormendy99} in modelling the surface brightness of the
bulge as a Sersic profile $I(R)=I_0\exp(-(R/R_n)^{1/n})$ with index
$n=2.19$, scale radius $R_n = 14\farcs0$ and central $V$-band surface
brightness $I_0=15.40$ mag.  Our kinematic model for the bulge is very
simple: it is non rotating and has a constant velocity dispersion
$\sigma_{\rm bulge}=120\kms$.  We add the contribution from this
simple bulge model to our models' predictions for the WFPC2 photometry
and the OASIS $V$ and~$\sigma$ maps.  We do not add it to the STIS
predictions; we assume that B05 have successfully removed the bulge
contribution from their STIS kinematics.

To model the contribution the young stars from P3 make to the $V$-band
light we include a another component having surface brightness
$\Sigma(X,Y)=\Sigma_3\exp(-R/R_0)$ in which $R^2=X^2+Y^2$ and the
scale length $R_0=0\farcs075$ (L12).  We follow B05 in assuming that
the kinematics extracted from the red spectra are unaffected by the
young stars; these stars affect only the WFPC2 photometry, not the
OASIS or STIS kinematics.

\subsection{Fitting the weights}

\label{sec:fit}
For given BH mass $M_\bullet$ and disc orientation
$(\theta_a,\theta_i,\theta_l)$, the model's prediction for any
observable $O_i$ can be be written as $\sum_jP_{ij}w_j$, in which
$w_j$ is the weight given to the $j^{\rm th}$ ring
(equ.~\ref{eq:ringsum}) and the matrix $P_{ij}$ gives the contribution
that the $j^{\rm th}$ ring makes to the $i^{\rm th}$ observable,
calculated using the method described in \S\ref{sec:ringobs} above. 
An observable $O_i$ can be the light within a ``superpixel'' (given by
WFPC2 photometry), a classical first- or second-order velocity moment
(OASIS kinematics) or a $0^{\rm th}$...$4^{\rm th}$-order modified
moment (STIS kinematics).  We do not include the zeroth-order moments
of the OASIS kinematics as these contain no additional information
over the WFPC2 photometry.

Having a vector of observables $(O_1,O_2,....)$ and associated
uncertainties $(\Delta_1,\Delta_2,\ldots)$, we use a non-negative
linear least squares algorithm \citep{lawson74} to find the
vector of non-negative weights $\vw$ that minimises
\begin{equation}
\chi^2 = \sum_i\left[\frac{O_i - \sum_jP_{ij}w_j}{\Delta_i}\right]^2.
\end{equation}
We then take this minimum value of $\chi^2$ as a measure of the
goodness of fit of the model with parameters
$(M_\bullet,\theta_l,\theta_i,\theta_a)$.

\section{Results}

Ideally, we would like to include all data sets (WFPC2, STIS, OASIS) in
our set of $O_i$ and fit simultaneously to them.  However, calculating
the contribution the $j^{\rm th}$ ring makes to a particular modified
moment is computationally intensive and a full study of the parameter
space spanned by $M_\bullet$, $\theta_l$, $\theta_i$ and $\theta_a$
that includes the STIS and OASIS data sets is not viable.
We therefore conduct our modelling in two stages.  We first fit models
to the WFPC2 photometry and the OASIS velocity distribution in order to
find the best set of orientation angles
$(\theta_a,\theta_i,\theta_l)$.  Then, having these angles, we fit to
the WFPC2 photometry and STIS kinematics to obtain our estimates of the
BH mass and the structure of the phase-space DF of the nucleus.  As a
final test of this model, we ``observe'' it at OASIS resolution and
compare it (by eye) to the real M31.

Before embarking on any of this model fitting, however, we first
confirm that our models are indeed able to reproduce some of PT03's
results.
\subsection{A test: reproducing PT03's model} 

\begin{figure*}\vspace{0.5cm}\includegraphics*[width=84mm, trim = 90 20 90 0]{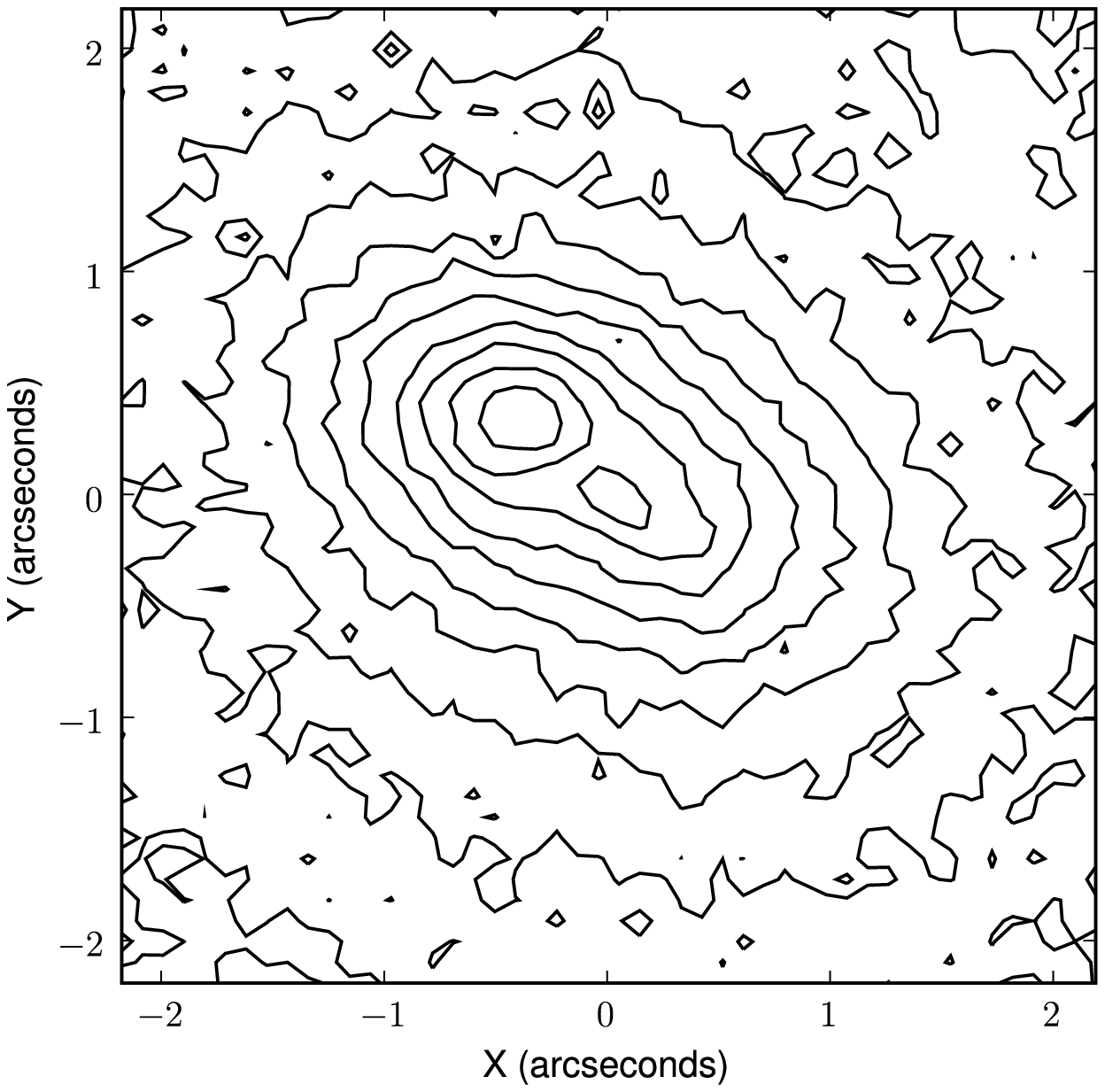}
\includegraphics*[width=84mm, trim = 90 20 90 -10]{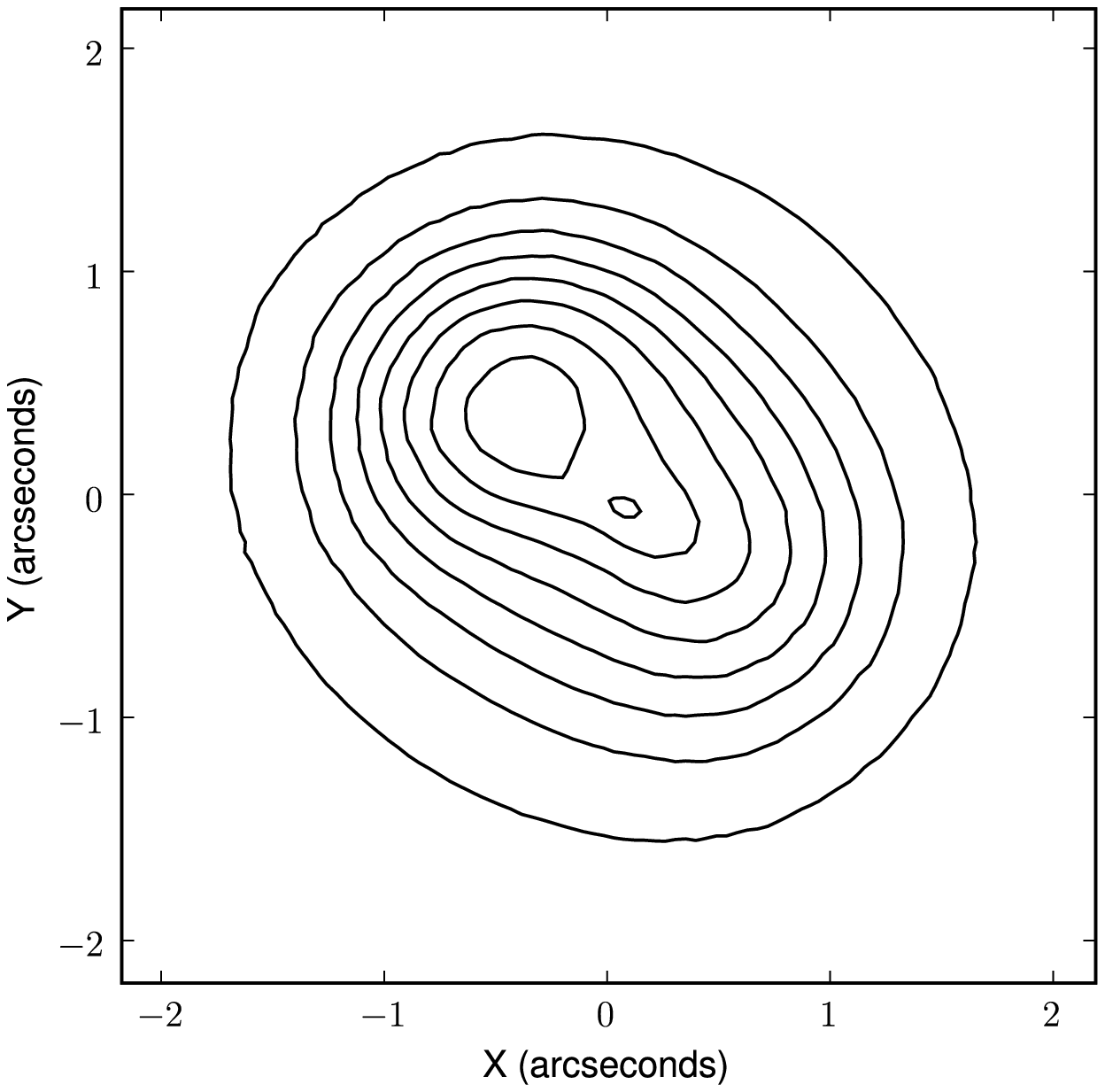}
\caption{Nuclear V-band surface brightness distribution.  The left panel shows the data and the
  right our reconstruction of Peiris \& Tremaine's (2003) non-aligned
  model.  Contours are at 0.25 mag intervals.
  Compare to figure~3 of PT03.}\vspace{0.5cm} \label{fig:pt03}
\end{figure*}

An immediate test of our modelling machinery is to reproduce some of
PT03's results by using a sum of rings~\eqref{eq:ring} that
approximates one of their DFs.  We focus here on their favoured
non-aligned model, but we found comparable results with the
poorer-fitting model that is forced to be aligned with the main M31
disc.  PT03's non-aligned models have DF
\begin{equation}
f(a, \ve, I) = g(a)\exp\left\{ - \frac{[\ve - \ve_m(a)]^2}{2\sigma_e^2}  \right\} \exp\left[ - \frac{I^2}{2\sigma_I(a)^2}  \right],
\end{equation}
where $\sigma_e = 0.307$.  The parametric form for the backbone eccentricities is given
by
\begin{equation}
\ve_m(a) = \alpha(a_e - a) \exp\left[-\frac{(a-a_g)^2}{2 w^2}\right]\hat\vx,
\label{eq:PT03em}
\end{equation}
with $\alpha = 0.197\,\rm pc^{-1}$, $a_e = 4.45$ pc, $a_g = 1.71$ pc and $w =
1.52$ pc.   For the dispersion in inclination
the form is
\begin{equation}
\sigma_I(a) = \sigma_I^0 \exp(-a/a_I)
\label{eq:PT03sigi}
\end{equation}
where $\sigma_I^0 = 24\fdg6$ and $a_I = 31.5$ pc.
The function $g(a)$ that sets the semi-major axis distribution
is
\begin{equation}
g(a) = \Sigma_0 \frac{a^{2} \exp(-a/a_0)}{1 + \exp[c_1(a-c_2)]} \; \frac{a^{1/2} }{2\upi^2 (G M_\bullet)^{3/2} \sigma_e^2 \sigma_I^2(a)},
\end{equation}
with $a_0 = 1.37$ pc, $M_\bullet=10.2\times10^8 M_\odot$, $c_1 = 4 $
pc$^{-1}$ and $c_2 = 4.24$ pc. We treat the overall normalisation $\Sigma_0$ as a free
parameter.

We approximate this DF by a sum of 720 rings of the form~\eqref{eq:ring} whose
semi-major axes $a_j$ are spaced logarithmically in radius from 0.1 pc
to 19 pc, with $\sigma_{a,j}=0.8a_j \Delta \log a$.  We set $\sigma_{e,j}=0.307$
for all rings and set the mean eccentricity
$\ve_j$ and dispersion in inclination $\sigma_{I,j}$ of the $j^{\rm
  th}$ ring by evaluating \eqref{eq:PT03em} and~\eqref{eq:PT03sigi} above at $a=a_j$.
The weights $w_j$ are set proportional to $a_jg(a_j)$, the factor of
$a_j$ coming from the fact that our rings are equispaced in $\log a$.

Figure~\ref{fig:pt03} shows the resulting model predictions for the
WFPC2 photometry when viewed at the same angles as PT03's best-fit
non-aligned model ($\theta_a = -34\fdg5$, $\theta_i = 54\fdg1$,
$\theta_l = -42.8\fdg2$) and including the contribution of the bulge
model from sec.~\ref{sec:bulge}.
Our reconstruction of the projected surface brightness in
their model agrees closely with their figure~3: the model produces a
nucleus that is broader than the observations and has an overly
extended flat profile at P1.

\subsection{Determining the orientation of the disc}

We conduct an exhaustive scan over the space of orientation angles
$(\theta_a,\theta_i,\theta_l)$ and black hole mass to obtain our best
guess for the orientation of the disc.  We first do this for a model
that is forced to be aligned ($\theta_i=77\fdg5$) with the
larger-scale M31 disc, before letting $\theta_i$ vary freely.
We include in these fits the 1123
OASIS $V$ data points as well as a broad field spanning $5\farcs824
\times 5\farcs824$ from the WFPC2 data: this utilises our varied
super-pixel scheme and contains 4096 data points. We do not include
the OASIS $\sigma$ maps in the fits, as it is dangerous to assume that
the measured $V^2+\sigma^2$ is a good estimate of the true second
moment of the LOSVD (figure~\ref{fig:typlosvd}).

\begin{figure}
\includegraphics*[width=\hsize, trim = 10 0 0 -30]{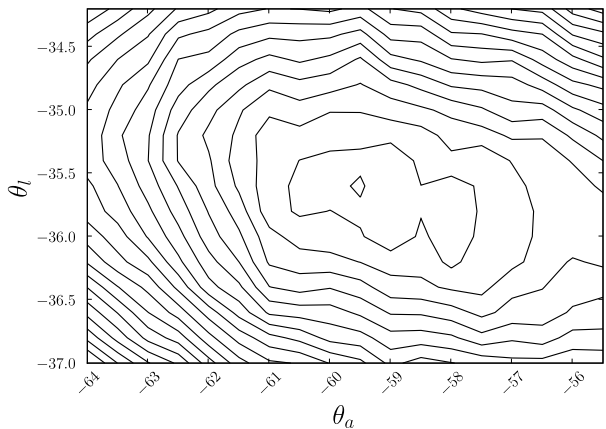}
\caption{ $\chi^2$ for $\theta_l$ vs. $\theta_a$ for the aligned
  model. Contours are spaced at $\Delta \chi^2 = 4$ intervals.}
\label{fig:aligned}
\end{figure}

For the aligned model we fix $\theta_i = 77\fdg5$ and take $\theta_a$
and $\theta_l$ in the intervals [$-90^\circ$,\;$0^\circ$] and
[$-55^\circ$,\;$-25^\circ$] with spacing $\Delta \theta = 2^\circ$.  A
contour plot depicting the shape of the $\chi^2$ goodness of fit for
the inner region of this range is shown in figure
\ref{fig:aligned}. As expected the aligned model constrains $\theta_l$
tightly but there is a weak dependence on $\theta_a$, with the best
fitting model falling at $\theta_a = -60^\circ$ and $\theta_l =
-35^\circ$.

\begin{figure*} 
\includegraphics*[width=58mm, trim = 10 10 0 0]{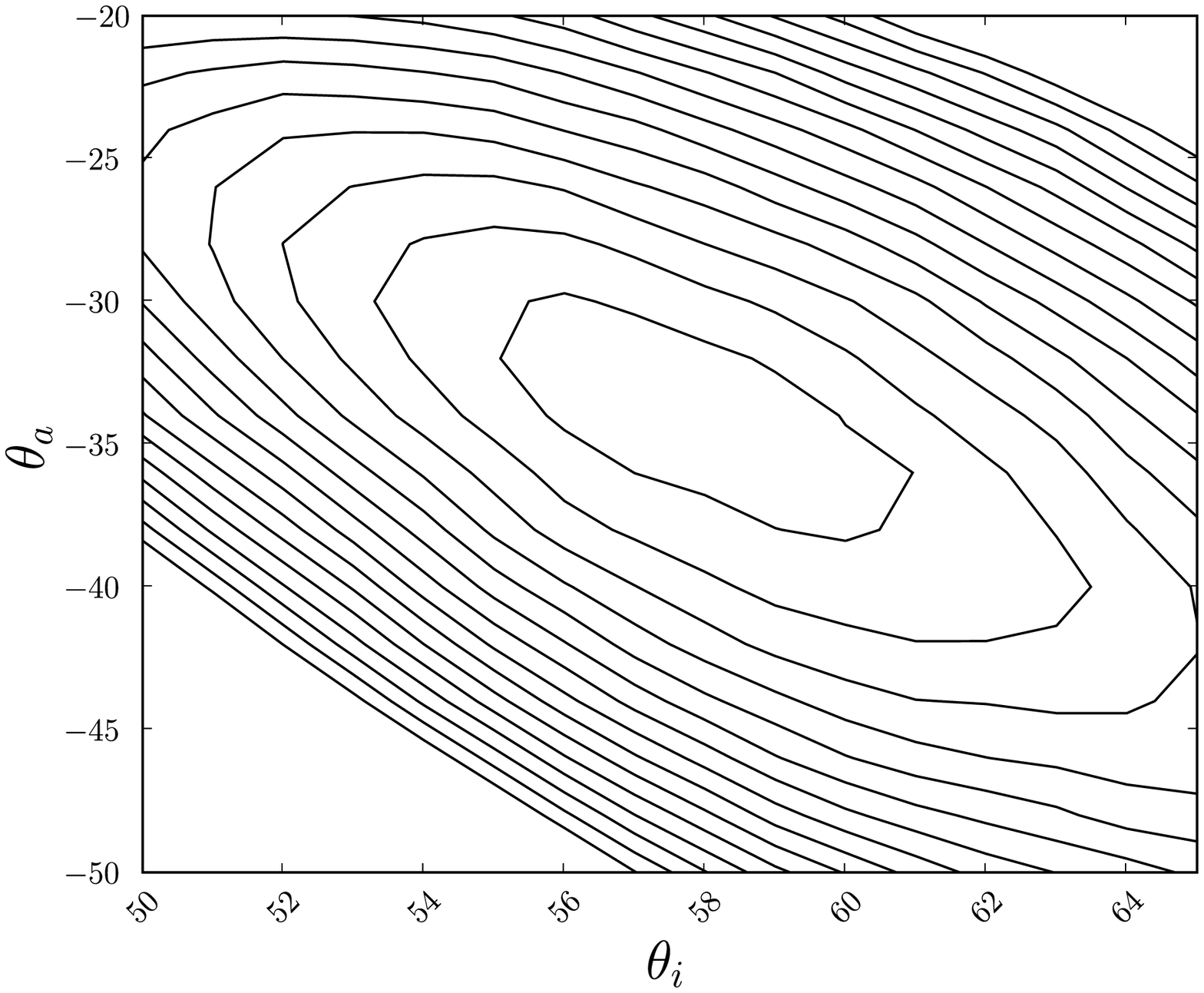}
\includegraphics*[width=58mm, trim = 10 10 0 0]{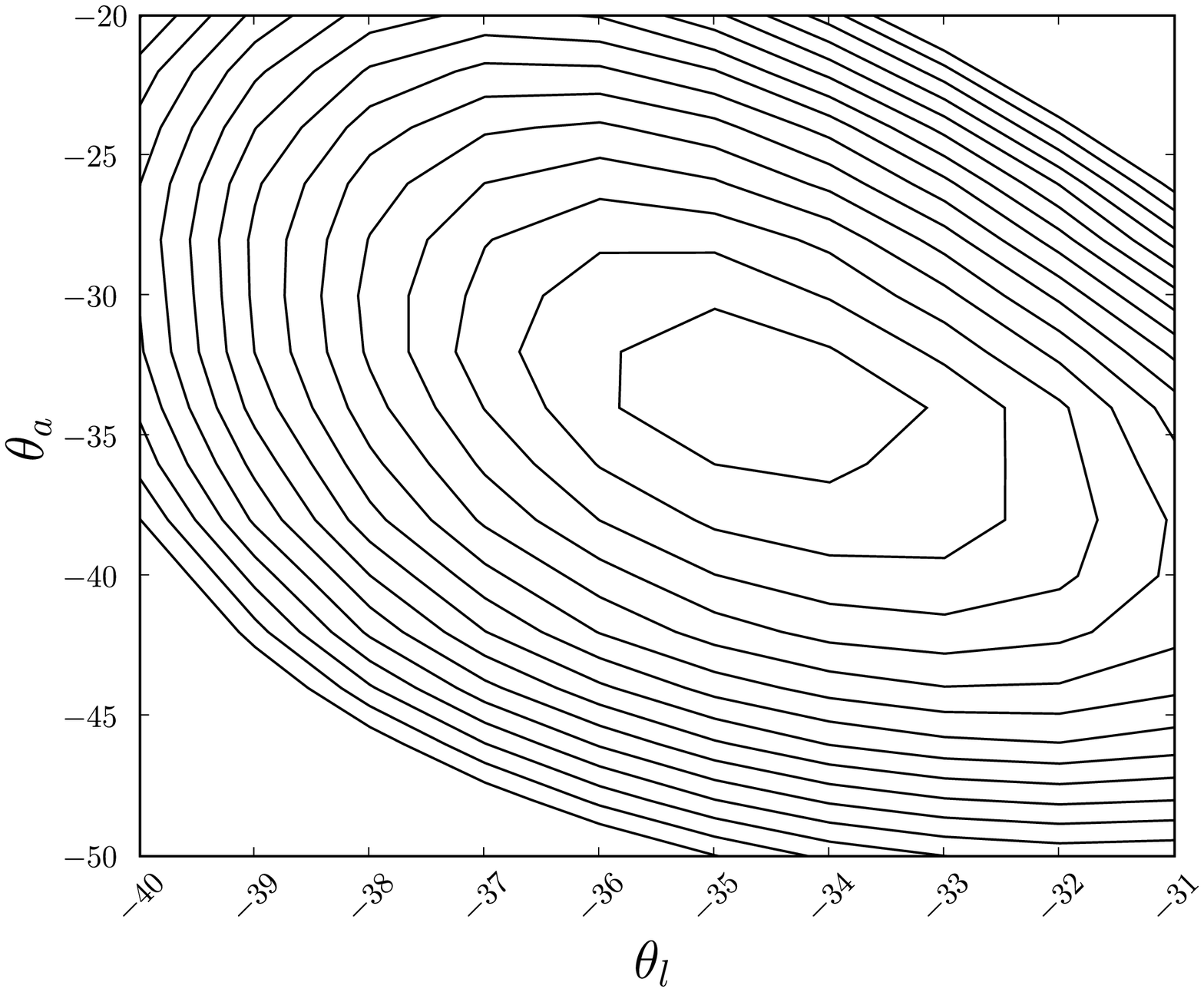}
\includegraphics*[width=58mm, trim = 10 10 0 0]{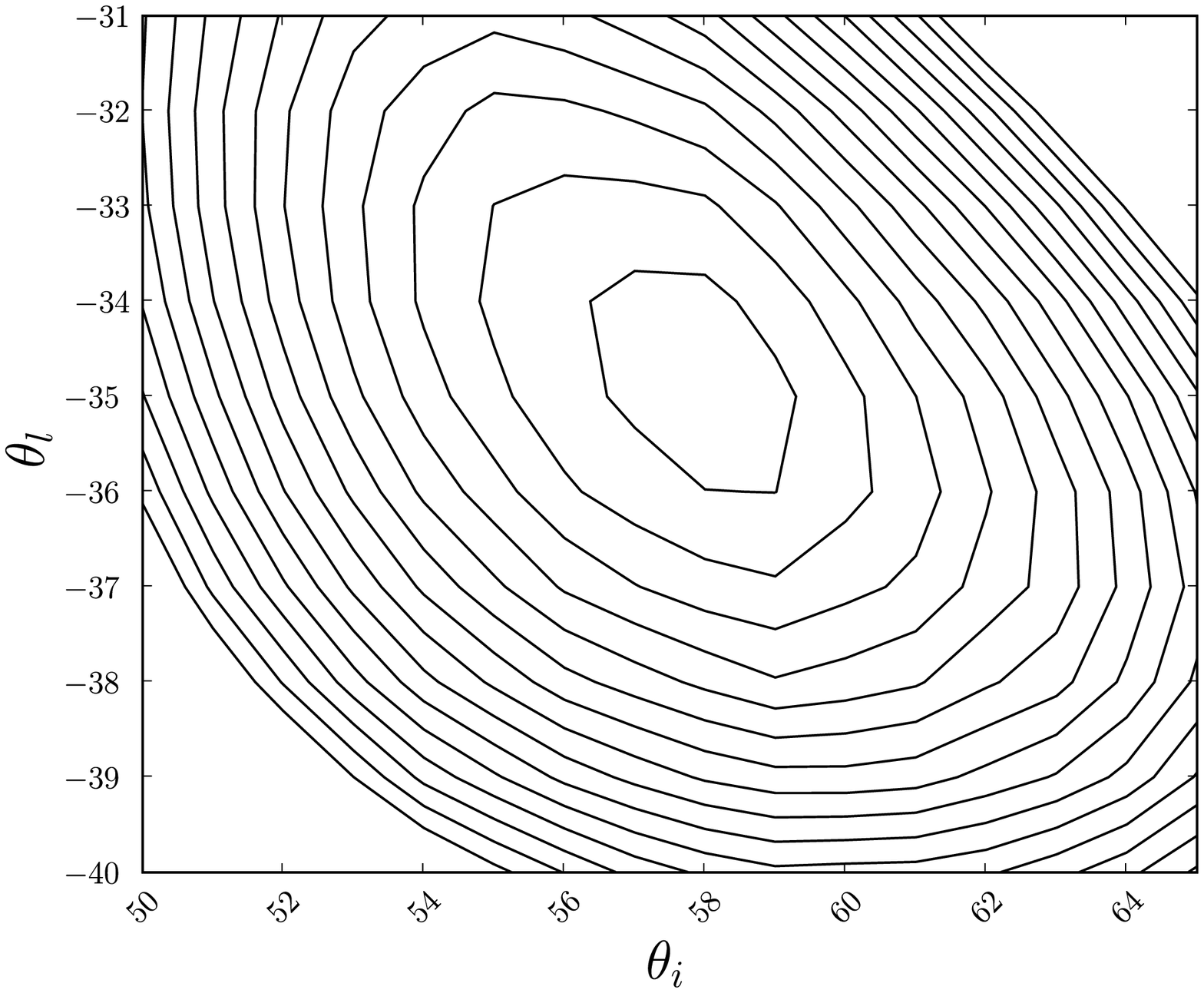}
\caption{Contour maps of $\chi^2$ for the slice through the best
  fitting model.  From left to right: $\theta_i$ vs. $\theta_a$ for
  $\theta_l = -35^\circ$; $\theta_l$ vs. $\theta_a$ for
  $\theta_i=57^\circ$; $\theta_i$ vs. $\theta_l$ for
  $\theta_a=-34^\circ$. Contours are spaced by $\Delta \chi^2 =
  25$.} \label{fig:contours}
\end{figure*}
\begin{figure*} 
\includegraphics*[width=58mm, trim = 10 10 0 0]{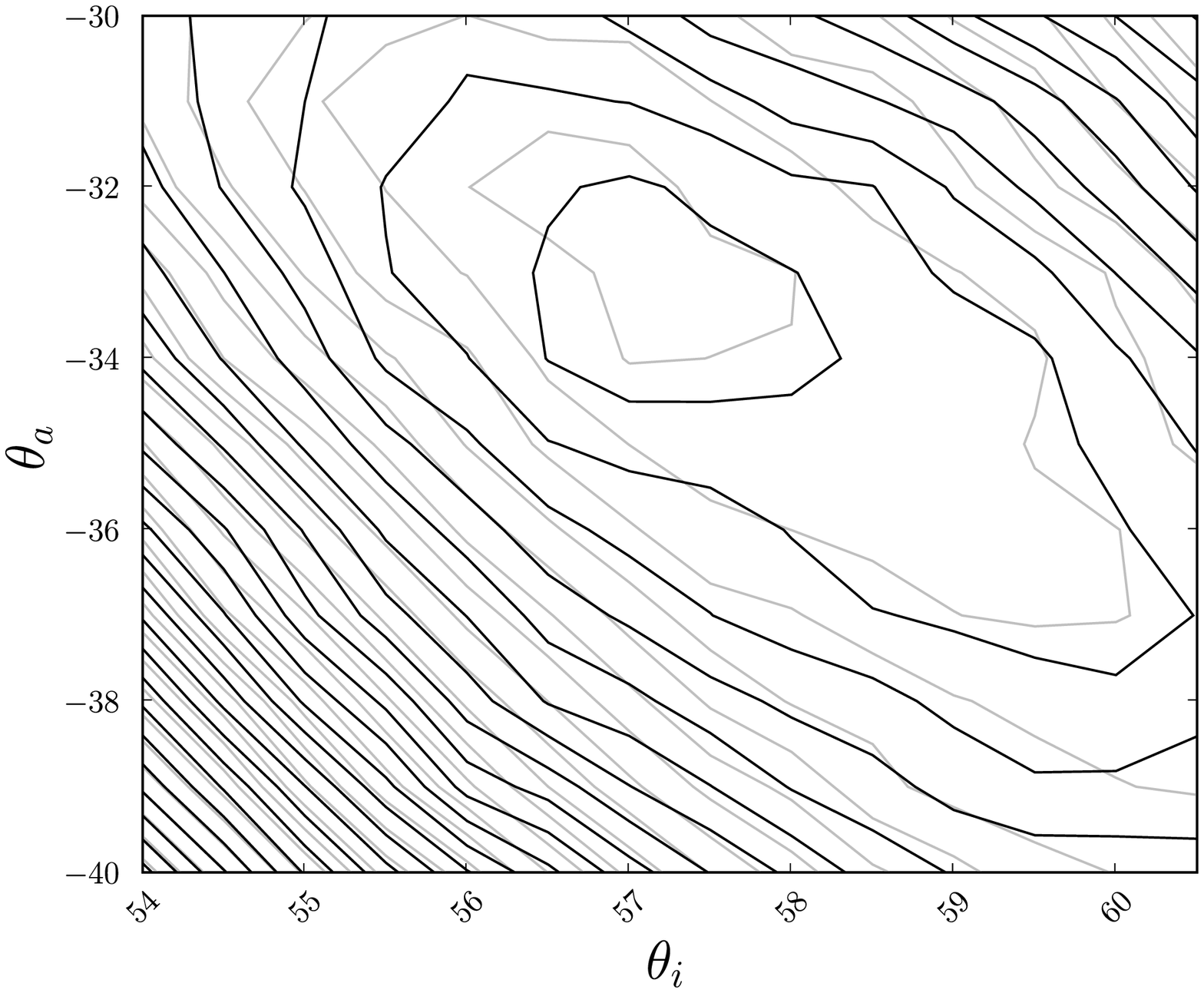}
\includegraphics*[width=58mm, trim = 10 10 0 0]{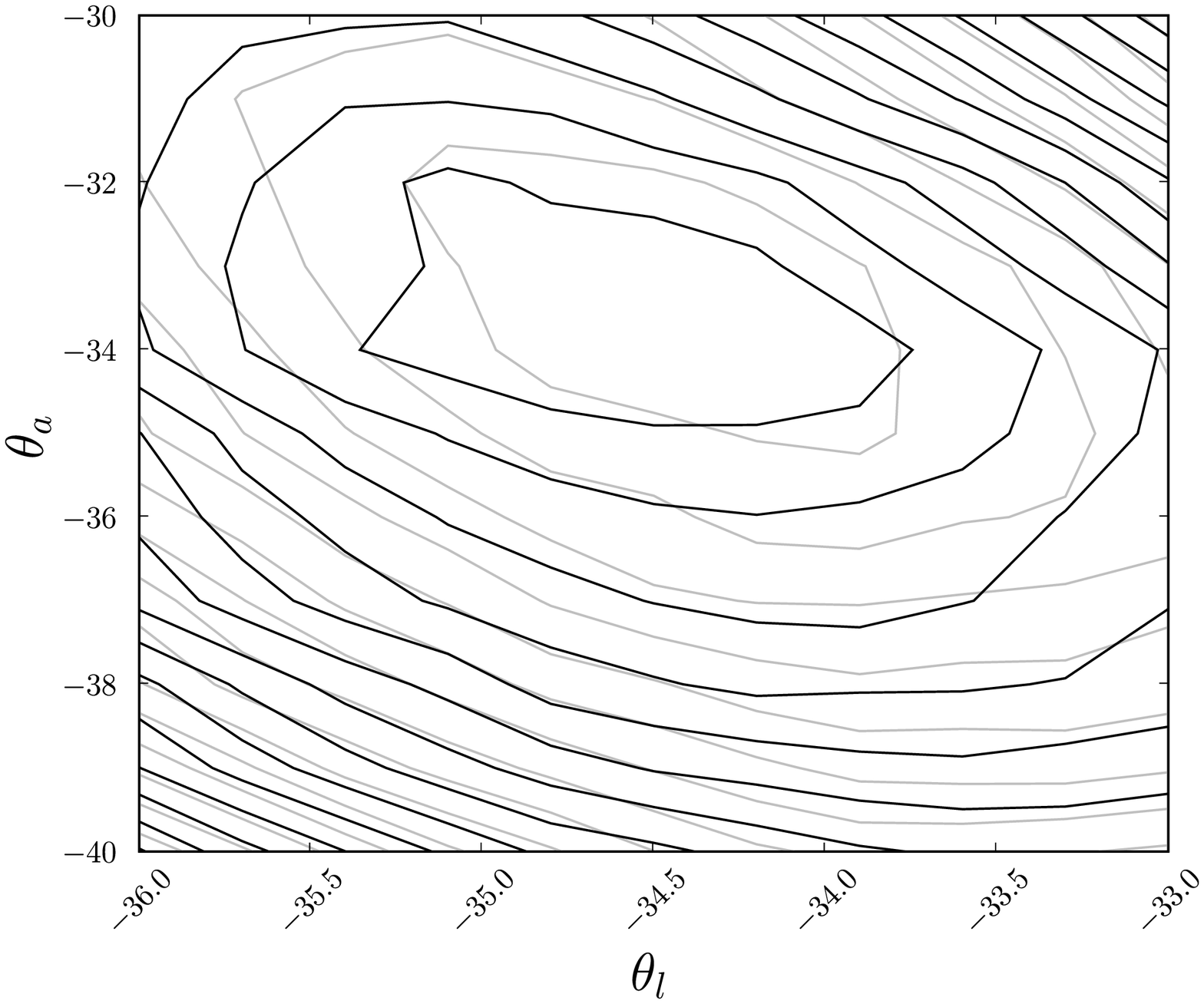}
\includegraphics*[width=58mm, trim = 10 10 0 0]{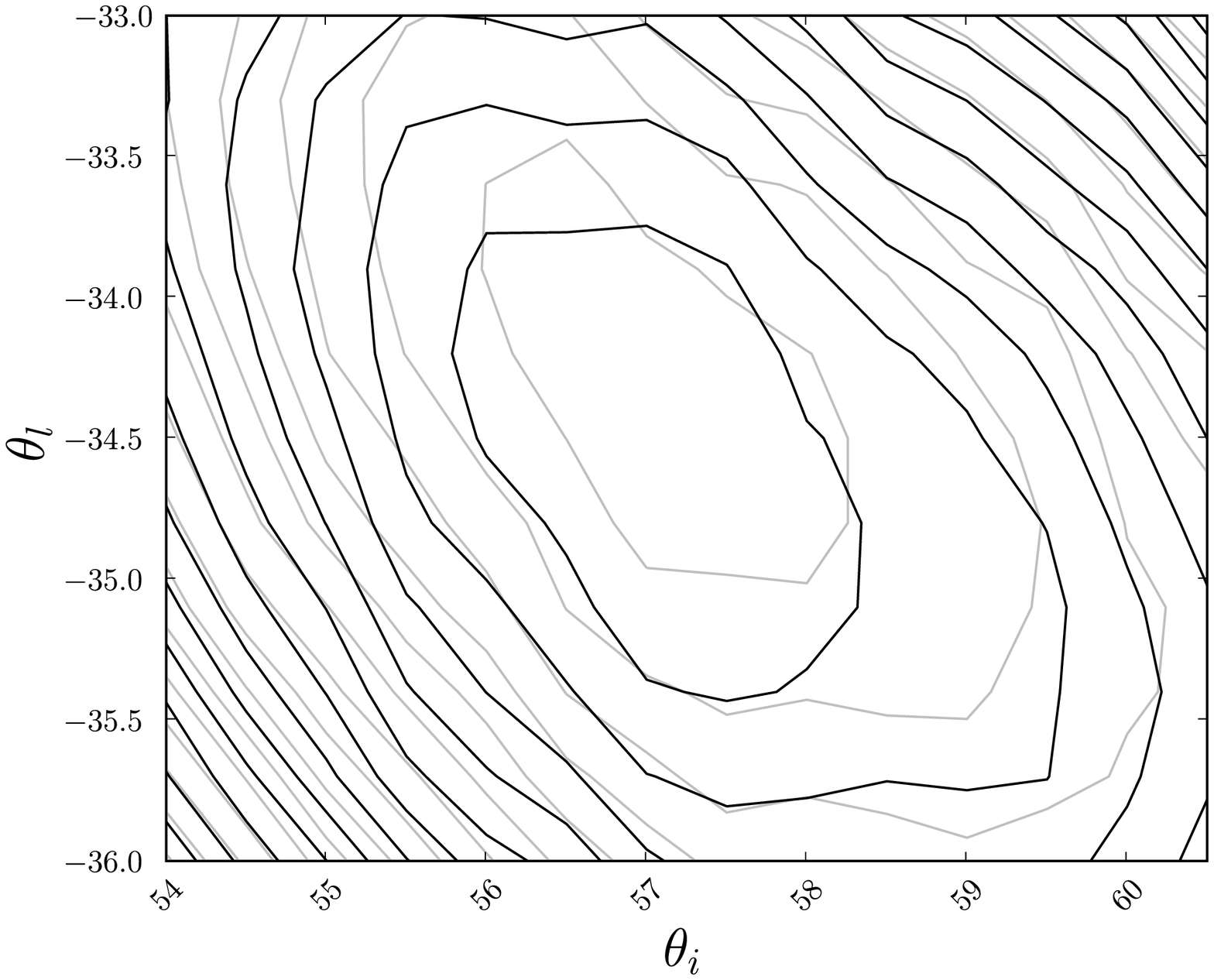}
\caption{Zoomed in contour maps. From left to right:
 $\theta_i$ vs. $\theta_a$; $\theta_l$ vs. $\theta_a$; $\theta_i$
 vs. $\theta_l$. Contours are spaced by $\Delta \chi^2 = 7.5$.
 The black contours show the same distribution as figure \ref{fig:contours}. The 
 grey contours show an alternative realisation of the distribution function to 
 illustrate the effect of shot noise.}
\label{fig:zoomed}
\end{figure*}

The non-aligned model looks at $\theta_a$ from $-58^\circ$ to
$-20^\circ$, $\theta_i$ from $50^\circ$ to $69^\circ$ and $\theta_l$
from $-39^\circ$ to $-29^\circ$, with $\Delta \theta_a = 2^\circ$ and
$\Delta \theta_l = \Delta \theta_i = 1^\circ$. The best fit appears at
$\theta_a = -34^\circ$, $\theta_i = 57^\circ$, $\theta_l = -35^\circ$
for a black hole mass $M_\bullet \simeq 1.25\times 10^8 M_\odot$. 
The shape of the $\chi^2$ goodness of fit for the non-aligned model is shown in figure ~\ref{fig:contours}. The presented scan is for
a single random distribution of stars projected with different angles:
scans using distributions drawn from a different random seed
found broadly similar results with slight variations (figure~\ref{fig:zoomed}). This presents issues with determining a single best
fitting set of angles. Using a much larger ($10^8$) number of stars
per disc in the region around the best fit angles informed our final
selection of specific angles, however even then the random seed
affected results. While in principle the angles could be determined
more accurately, in practice shot noise from the model (and also the
finite number of stars in the real nucleus!) limits the accuracy in
each angle to the order of $\approx1^\circ$.  Our final selection of angles was
determined from the projection of the likelihood $\exp(-\frac12\chi^2)$ onto
each of the $\theta_a$, $\theta_i$ and $\theta_l$ axes.

We have experimented with allowing the bulge surface brightness $I_0$
and the central surface brightness $\Sigma_3$ of the young stellar
disc described in sec.~\ref{sec:bulge} to float our fitting procedure
by including them as additional ``weights'' $w_j$ in the model and
adding two additional columns to the projection matrix $P_{ij}$, but
we find that this makes little difference to our results.

\subsection{Fitting WFPC2 photometry and STIS kinematics}
\label{sec:gamma}

Having the orientation angles we now drop the OASIS $V$ maps and focus
on using WFPC2 photometry together with the STIS LOSVDs to further
constrain the model.  The region of the WFPC2 photometry we use is
restricted to an ellipse of semi-major axis $1\farcs6$ and axis ratio
0.6 centred on P2. This ellipse is just large enough to encompass all
STIS positions.  \updated{}{We rebin} \citet{lauer98}'s dithered image 4 by 4 into ``superpixels''
of side 0\farcs0455.  Our vector of observables $O_i$ consists of the
WFPC2 fluxes in all 2335 such superpixels that lie
within the ellipse, together with the 5
modified moments $(\tilde h_{i,0},...,\tilde h_{i,4})$ obtained from
$(\gamma_i,V_i,\sigma_i,h_{3,i},h_{4_i})$ for each of the
$i=1...22$ LOSVDs measured by \cite{bender05} using the procedure
described earlier in section~\ref{sec:stismodmom}.  

\subsubsection{Reconstruction of the $\gamma(R)$ profile}
\begin{figure}
  \centering
  \includegraphics[width=\hsize, trim = 20 20 20 20]{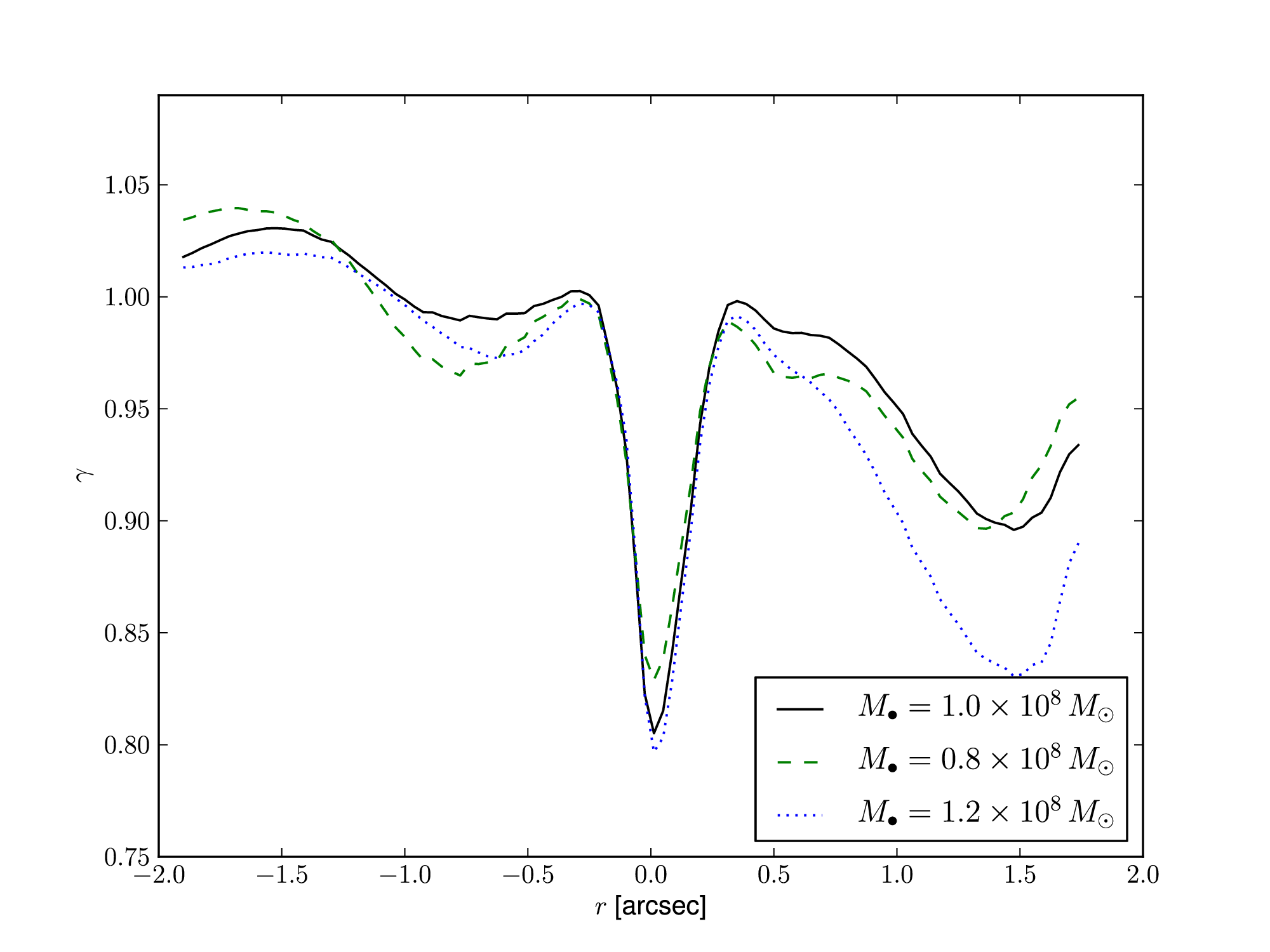}
  \caption{The Gauss--Hermite ``line-strength'' parameter
    $\gamma$ reconstructed along the STIS slit using the iterative
    procedure described in section~\ref{sec:gamma} for three different
    assumed black hole masses, as indicated on the legend.}
  \label{fig:gamma}
\end{figure}
The only complication in this is that we do not know the line
strengths $\gamma$ for any of our LOSVDs.  We do, however, expect
$\gamma$ to be reasonably close to one and so we first fit a model in
which all $\gamma_i=1$.  Then, having the weights $w_j$, we construct
a realization of this model and ``observe'' it convolved with the STIS
PSF~\eqref{eq:STISPSF}.  We fit Gauss--Hermite coefficients
$(\gamma,V,\sigma,h_3,h_4)$ to the model LOSVDs at each point along
the STIS slit, giving us a more informed estimate of how $\gamma$
varies along the slit.  We then repeat the whole fitting procedure,
replacing our original $\gamma_i=1$ guesses with values read off from
this reconstructed $\gamma(R)$ distribution.  We find that the
resulting model converges after only a couple of iterations of
this scheme.  Figure~\ref{fig:gamma} plots representative $\gamma$ profiles
obtained by this procedure for models with black hole masses $M_\bullet/10^8M_\odot=1.0$,
0.8 and 1.2.  This shows that $\gamma$ is significantly depressed
close to the black hole where the LOSVDs are least well described by
simple Gaussians.

\begin{figure*}
  \centering
  \includegraphics*[width=0.49\hsize, trim = 20 10 20 10]{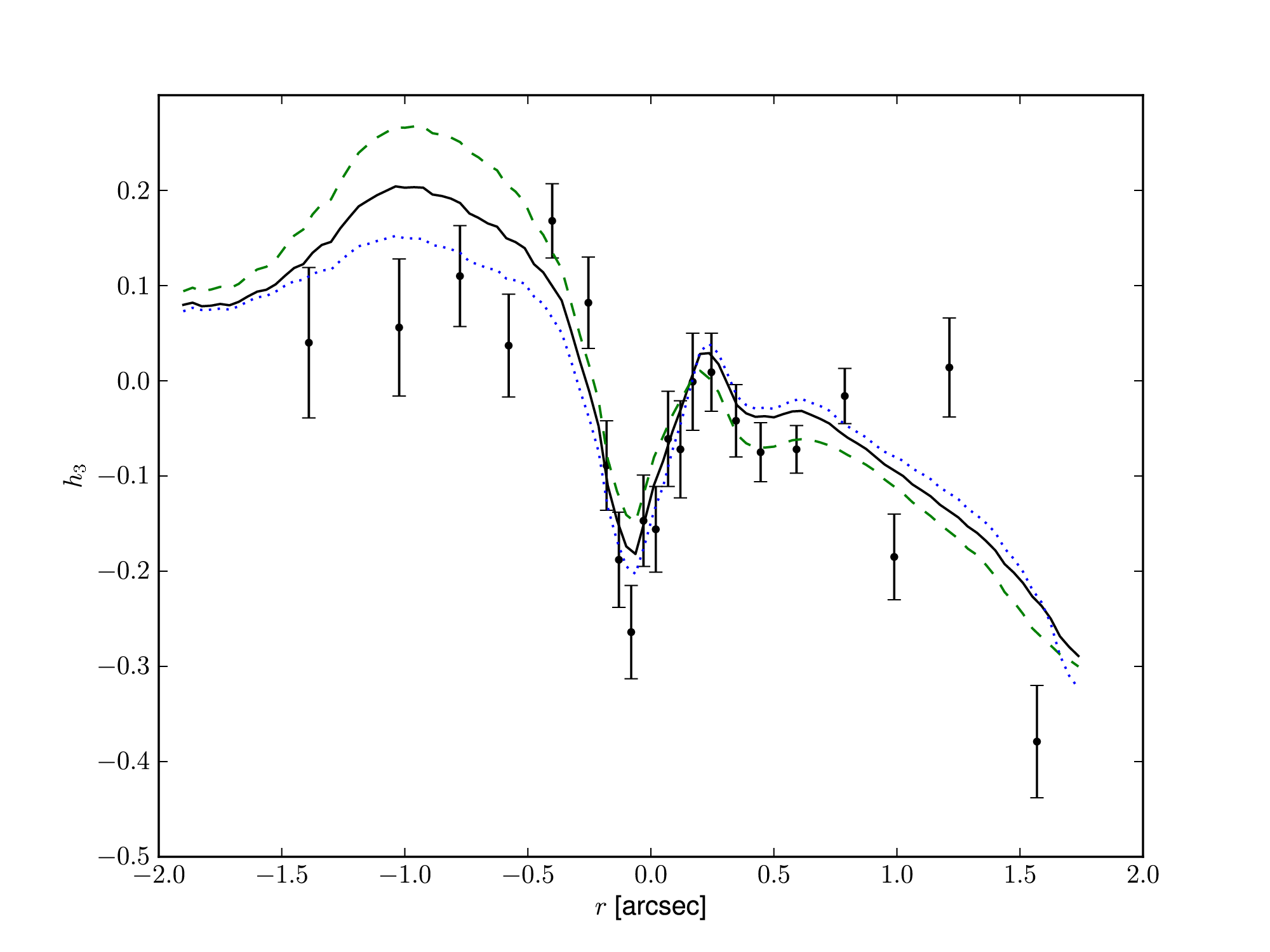}
  \includegraphics*[width=0.49\hsize, trim = 20 10 20 10]{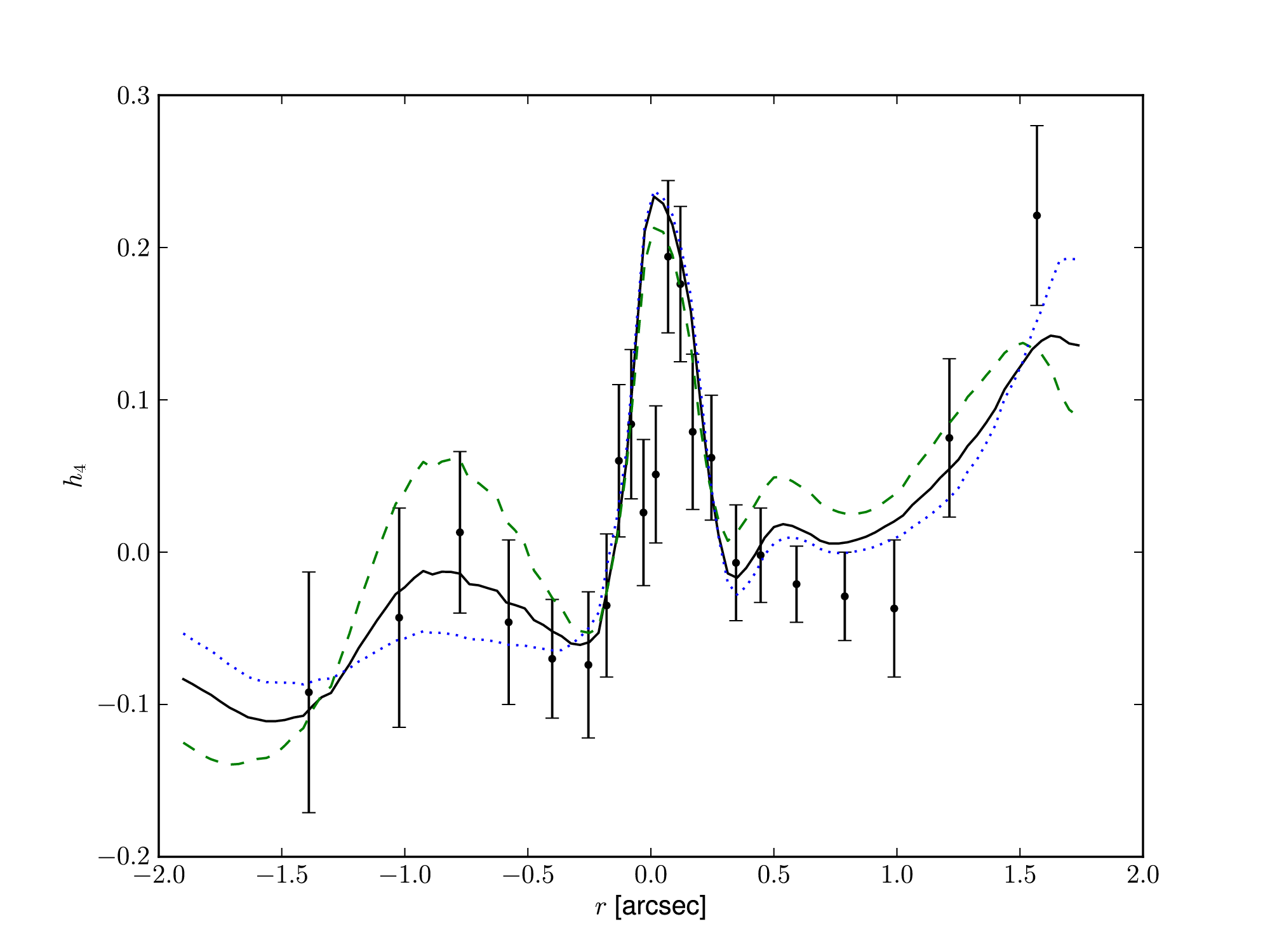}
  
  \includegraphics*[width=0.49\hsize, trim = 20 10 20 10]{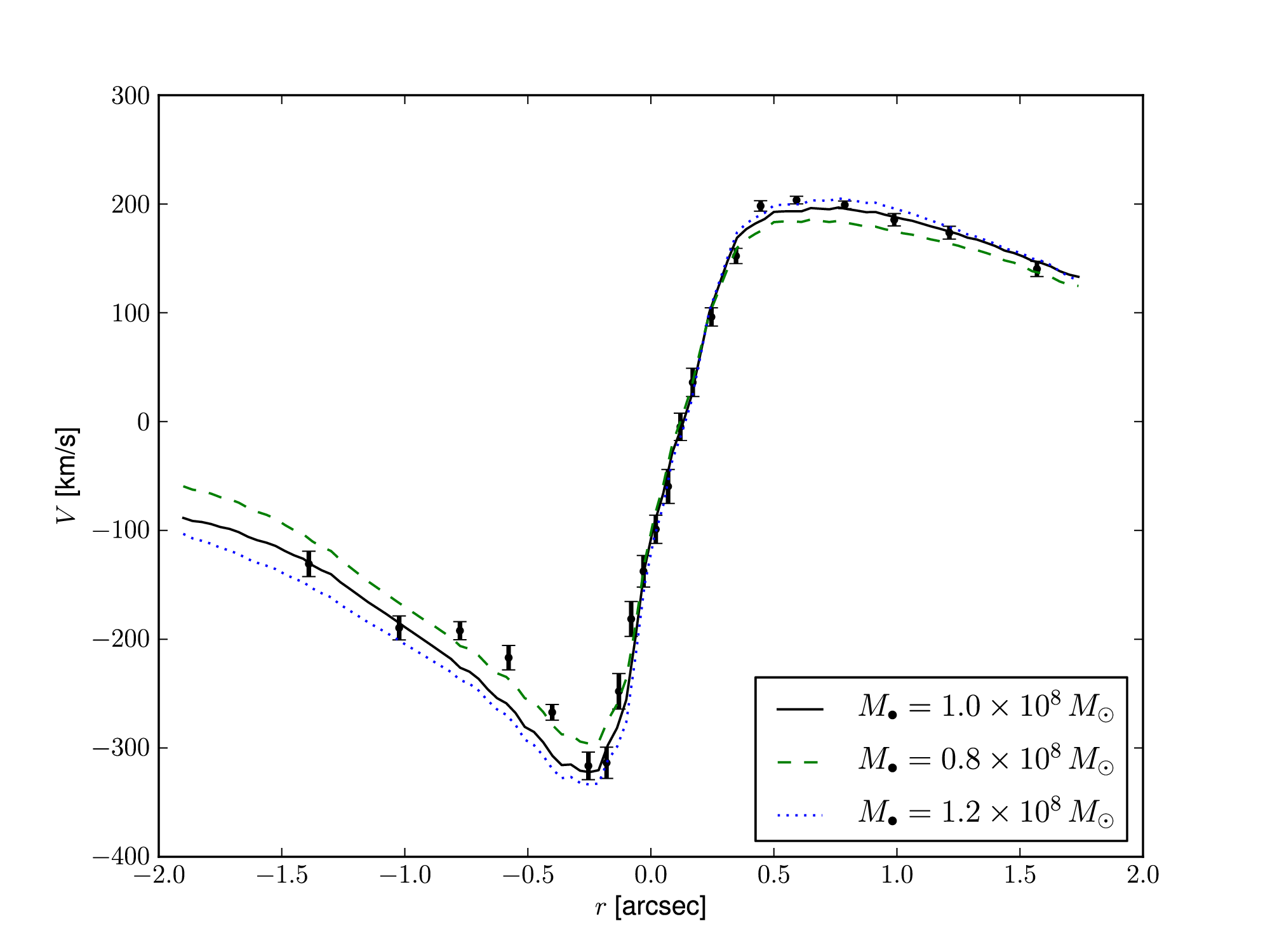}
  \includegraphics*[width=0.49\hsize, trim = 20 10 20 10]{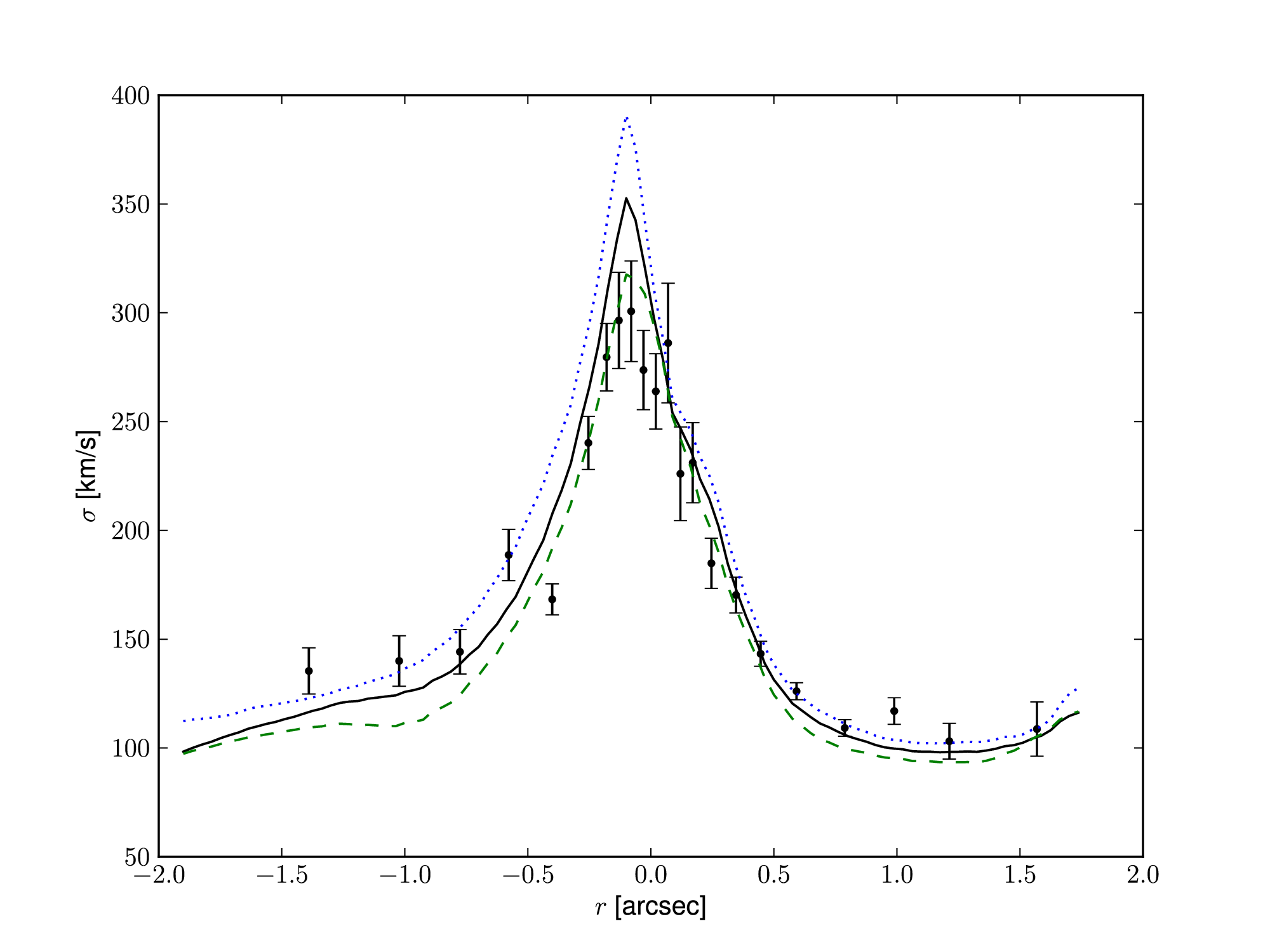}
  \caption{Gauss--Hermite coefficients parametrising the LOSVDs along
    the STIS slit measured by B05 (points) together with our model
    fits for three different assumed black hole masses (curves).}
  \label{fig:stis_gh}
\end{figure*}

\subsubsection{Comparison of best-fit model against observations}
The formal $\chi^2$ of the model with $M_\bullet=1.0\times10^8M_\odot$
is 782.  To put this in perspective, the vector of observables $O_i$
that the model fits has 2445 elements: 2335 WFPC2 fluxes plus
$5\times22$ modified moments.  For comparison, the model with
$M_\bullet=0.8\times10^8\,M_\odot$ has $\chi^2=825$, while the model
with $M_\bullet=1.2\times10^8\,M_\odot$ has $\chi^2=859$.
Figure~\ref{fig:stis_gh} shows the Gauss--Hermite coefficients of the
reconstructed models along the STIS slit.  The agreement with B05's
observed kinematics is good, 
but there some features the best
$M_\bullet=1.0\times10^8\,M_\odot$ model cannot reproduce: the model
does not fit 
the detailed shape of the $V(R)$ and $h_3(R)$ profiles  between
$R=-0\farcs9$ and $0\farcs3$ well and it predicts a central $\sigma(R)$
that is slightly too high.

\begin{figure}
  \centering
  \includegraphics*[width=\hsize, trim = 30 30 30 30]{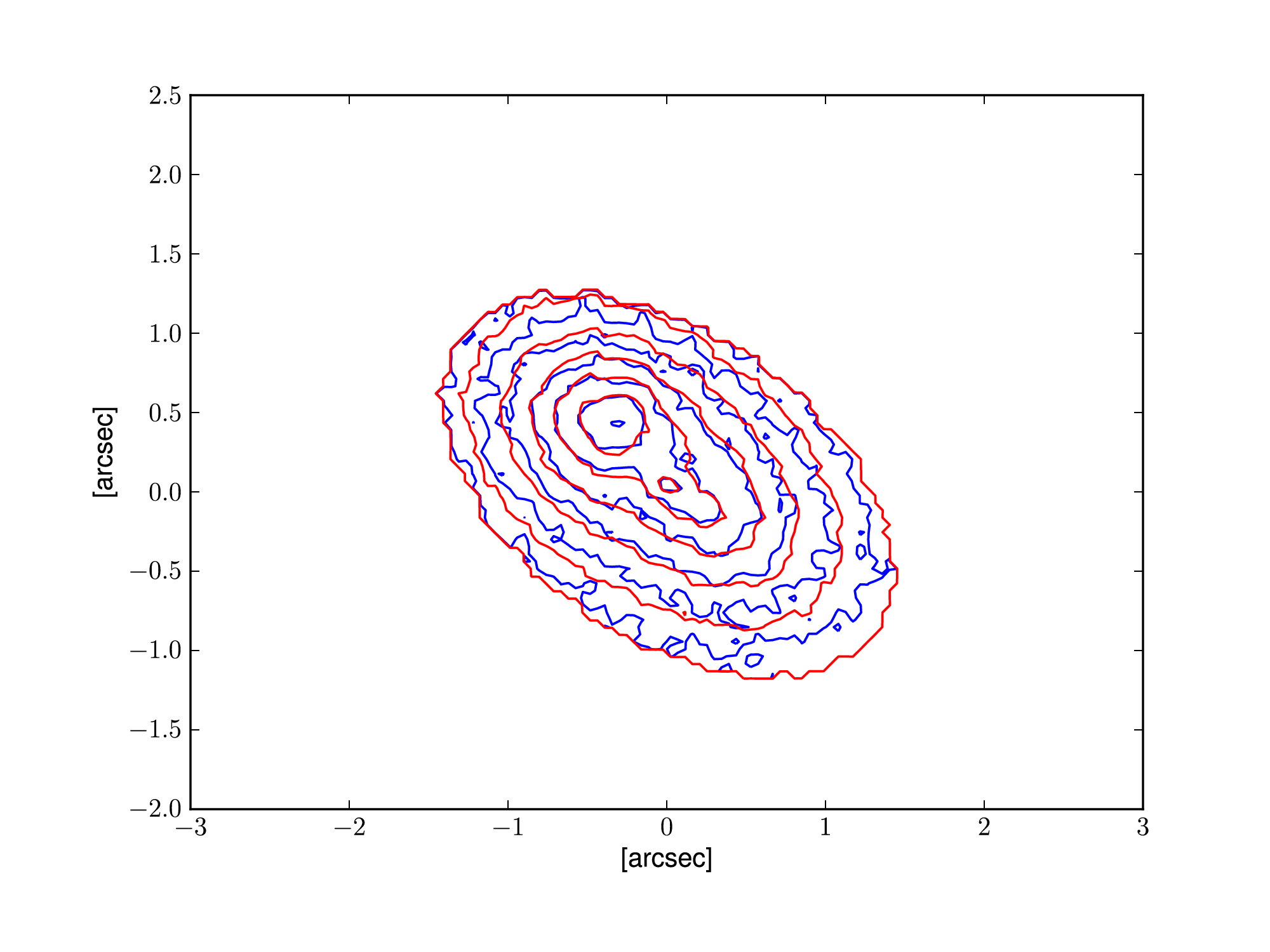}
    \caption{Observed $V$-band WFPC2 image (blue) with the fit from
      our $M_\bullet=1.0\times10^8M_\odot$ overlaid
      on top (red).  Contours are spaced at 0.3 magnitude intervals.}
    \label{fig:wfpc}
\end{figure}

Figure~\ref{fig:wfpc} shows how well the
$M_\bullet=1.0\times10^8\,M_\odot$ model fits to the WFPC2 photometry.
Results for the other two black hole masses are similar.  The
agreement is good in the central regions, but beyond about 1 arcsec
from P2 the model's surface brightness profile falls off too slowly
compared to the observations.  One possible explanation for this is
that our model is simply too coarse; the thickness of each of the
$n_a=20$ rings (equ.~\ref{eq:ring}) in these models is
$\sigma_a=0.23a$, which sets the models' characteristic radial spatial
resolution.  Another is that our model for the contribution of the
bulge light (sec.~\ref{sec:bulge}) may be wrong within the innermost
couple of arcsec.

Based on these comparisons, we interpret the relatively low values of
$\chi^2$ of our models not as a indication of the outstanding quality
of our model fits, but instead as a sign that the treatment of the
observational uncertainties -- particularly of the WFPC2 photometry --
could be improved.  Nevertheless, we believe that the reader will
agree that simple ``chi-by-eye'' tests indicate that our models
produce the best fits to date of the M31 eccentric disc system.  Of
course, this is to be expected given that we have $>700$ free
parameters to play with, which is at least an order of magnitude more
than most previous models of M31's nucleus.

\begin{figure*}
  \centering
  \includegraphics*[width=0.49\hsize, trim = 20 40 40 20]{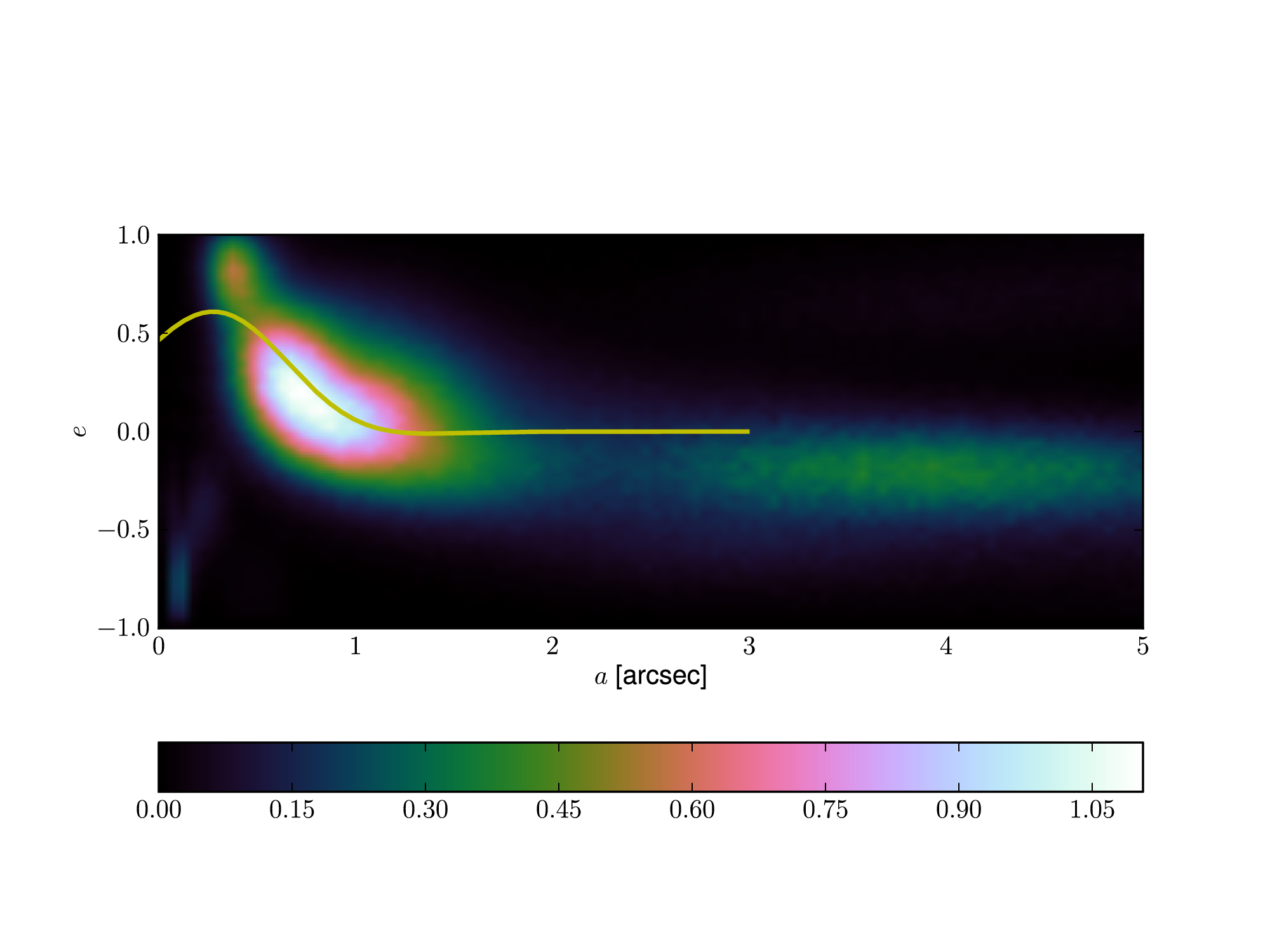}
  \includegraphics*[width=0.49\hsize, trim = 20 40 40 20]{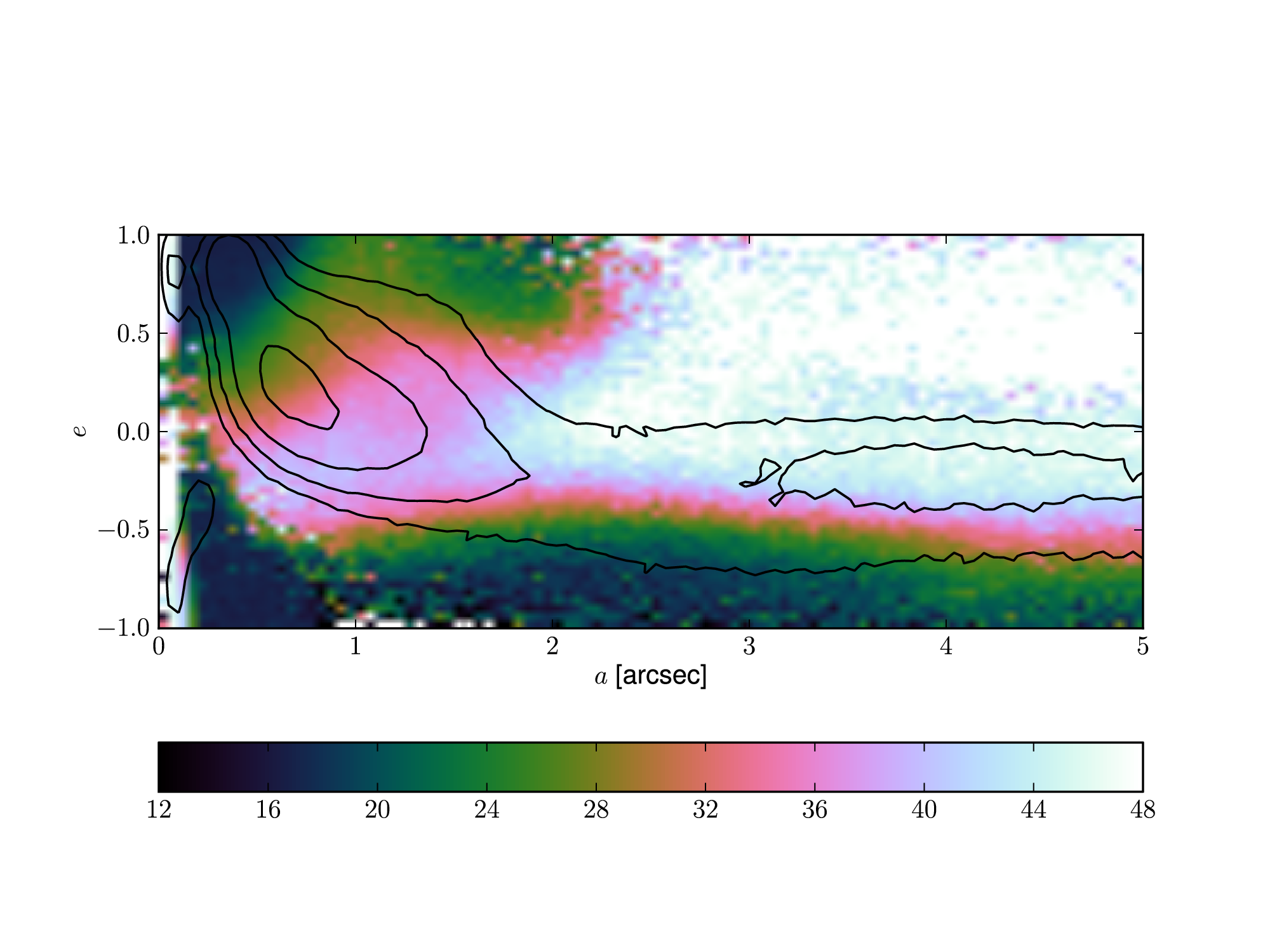}
    \caption{The distribution of orbits in our best-fit model with
      $M_\bullet=1.0\times10^8\,M_\odot$.  Left panel:
      the projected DF $f(a,e_x)$ \updated{}{obtained from the full DF $f(a,e_x,e_y,I)$ by averaging
        over $e_y$ and~$I$.}
The yellow curve plots the $a(e_x)$ profile of
      PT03's best-fit model for comparison.  Right panel: 
      RMS inclination angle (in degrees) as a function of $(a,e_x)$
      \updated{}{obtained from $f(a,e_x,e_y,I)$ by averaging over
        $e_y$}.
      The $f(a,e_x)$
      distribution from the left panel is overlaid as contours.}
    \label{fig:DF}
\end{figure*}
 
\begin{figure}
  \centering
  \includegraphics*[width=\hsize, trim = 20 10 40 20]{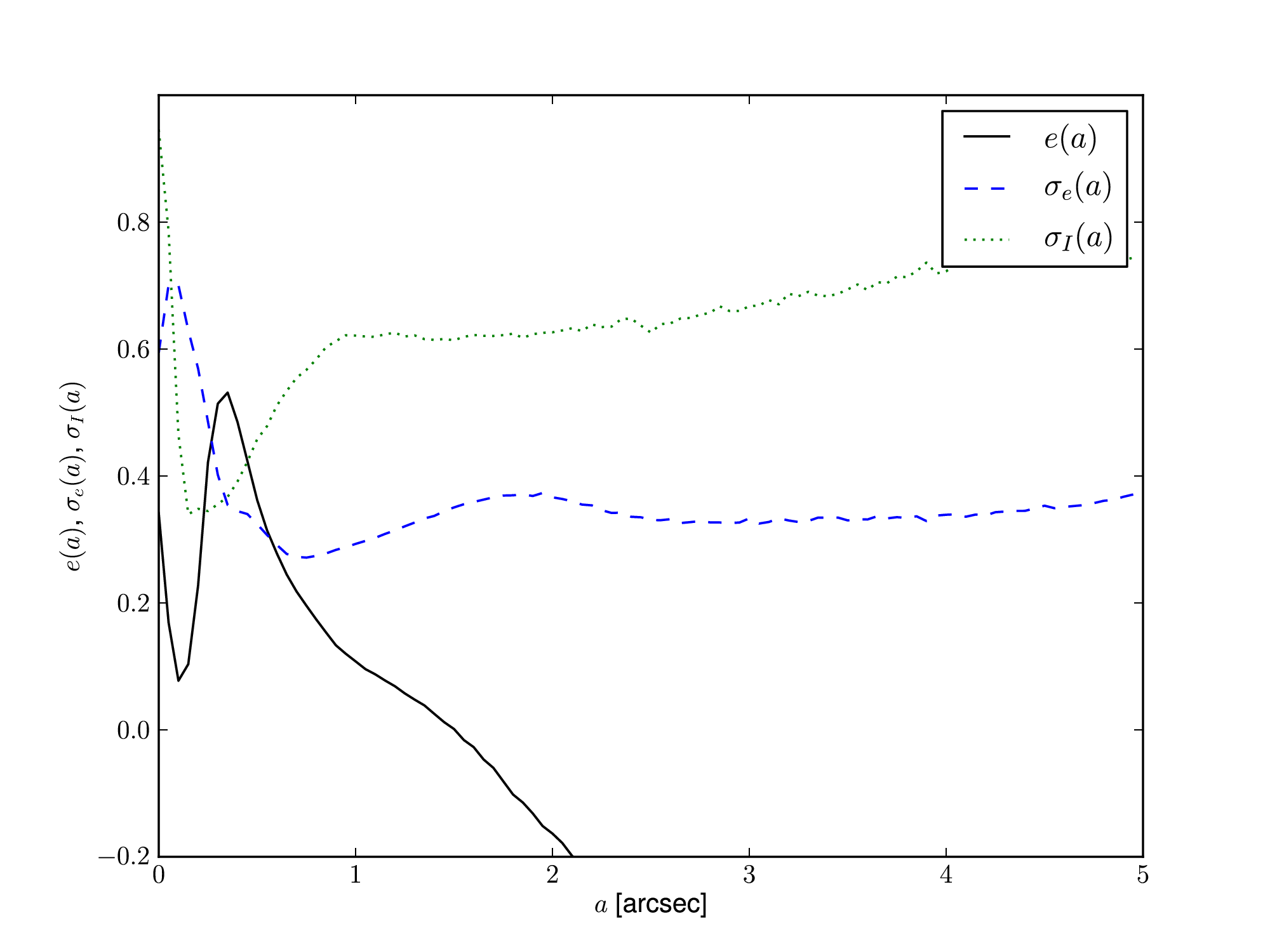}
  \caption{Mean eccentricity $e_x(a)$ and dispersions in eccentricity
    $\sigma_{e,x}(a)$ and inclination $\sigma_I(a)$ as a function of
    semi-major axis~$a$.  Unlike figure~\ref{fig:DF} the dispersion
    $\sigma_I$ in this plot is given in radians, not degrees.}
    \label{fig:DFsig}
\end{figure}
\subsubsection{DF of best-fitting model}
What can we learn from all these free parameters, specifically the
orbit weights $w_j$?  Figure~\ref{fig:DF} shows two views of the DF
\updated{}{$f(a, e_x, e_y, I)$} of our best-fit model.  This model has a strong negative
eccentricity gradient between 0\farcs5 and 1\farcs2 (corresponding to
P1), which is very similar to PT03's best-fit model; even the
dispersion in eccentricity (figure~\ref{fig:DFsig}) is similar to their
value of $0.307$.  There are three important differences between our
model and theirs, however.

\begin{enumerate}
\item Beyond 1\farcs2 the mean eccentricity
in our models becomes negative, meaning that the rings become mildly
antialigned.  We find that the strength of this feature depends on the
details of our bulge model and so it is hard to judge its
significance, but we note that just such a feature was predicted by
\cite{statler99} in his analysis of thin, self-gravitating discs.
\item Whereas PT03's parametrised model had an exponentially declining
$\sigma_I$ profile, our model fits a $\sigma_I$ profile that {\it
  increases} with radius for $a>0\farcs15$.  This last point is
qualitatively consistent with the predictions of collisional models of
disc evolution \citep{stewart00,peiris03}.  The detailed agreement is
not so good though: whereas the collisional models predict
$\sigma_I/\sigma_e\simeq0.5$, our models fit
$\sigma_I/\sigma_e\simeq2$.  It is not immediately clear, however, how
far one can apply these calculations that assume almost circular $e=0$
discs to the strongly eccentric disc in M31.
  \item $f(a)$ in our model falls steeply towards the centre from
$a=0\farcs5$ to $a\simeq0\farcs15$, inside which there is antialigned
($e_x\simeq-0.5$), fat ($\sigma_I=48^\circ$) distribution of orbits.
Recall that the range of $\sigma_I$ reproducible by our chosen sample
of rings is from $12^\circ$ to $48^\circ$.
This distinctive change in the distribution of the red stars is almost
cospatial with the young, A-star population that make up P3.
\end{enumerate}

\subsubsection{Other projections of the best-fit model}
\begin{figure*}
  \includegraphics*[width=0.49\hsize, trim = 50 20 20 20]{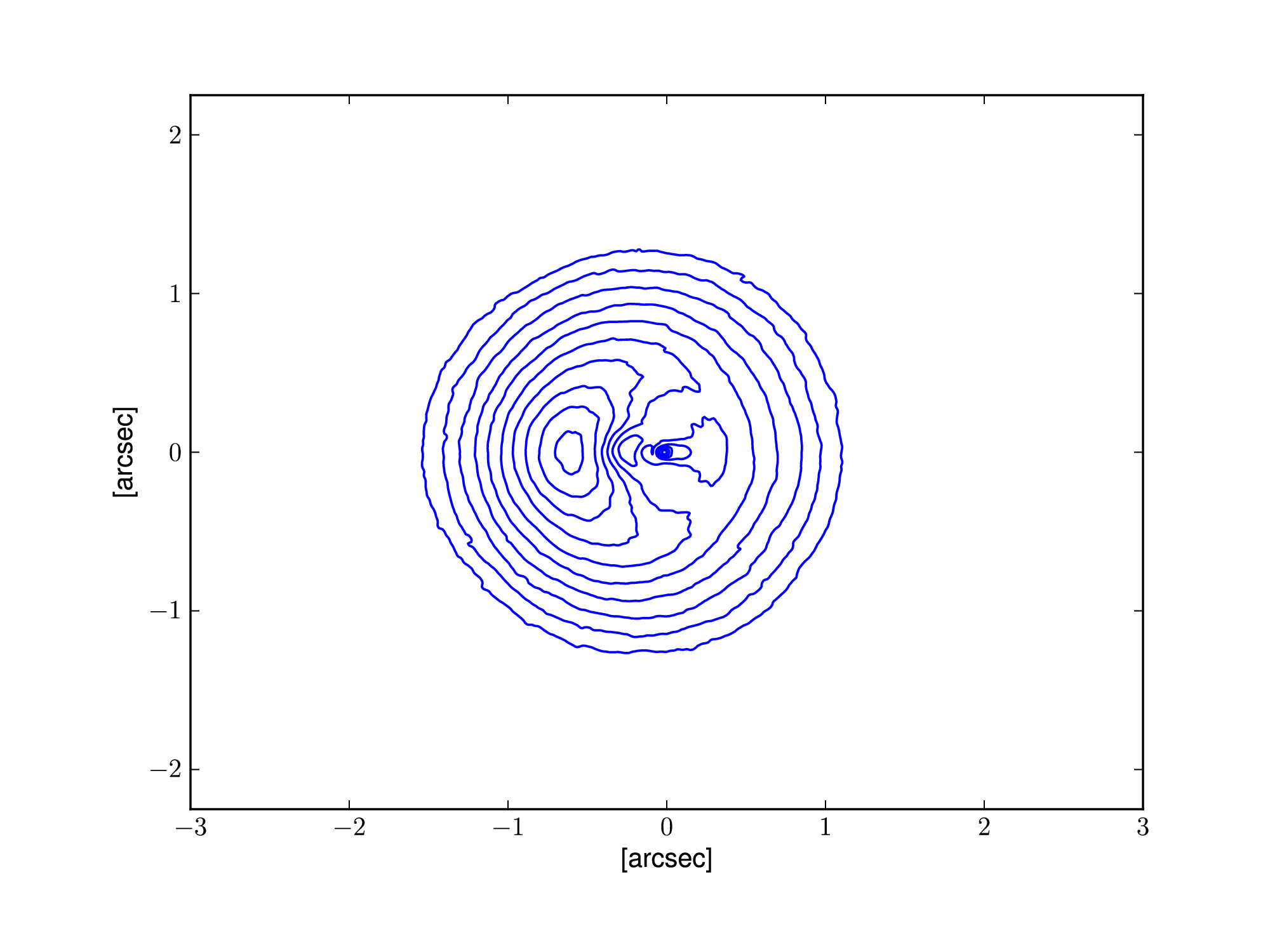}
  \includegraphics*[width=0.49\hsize, trim = 50 20 20 20]{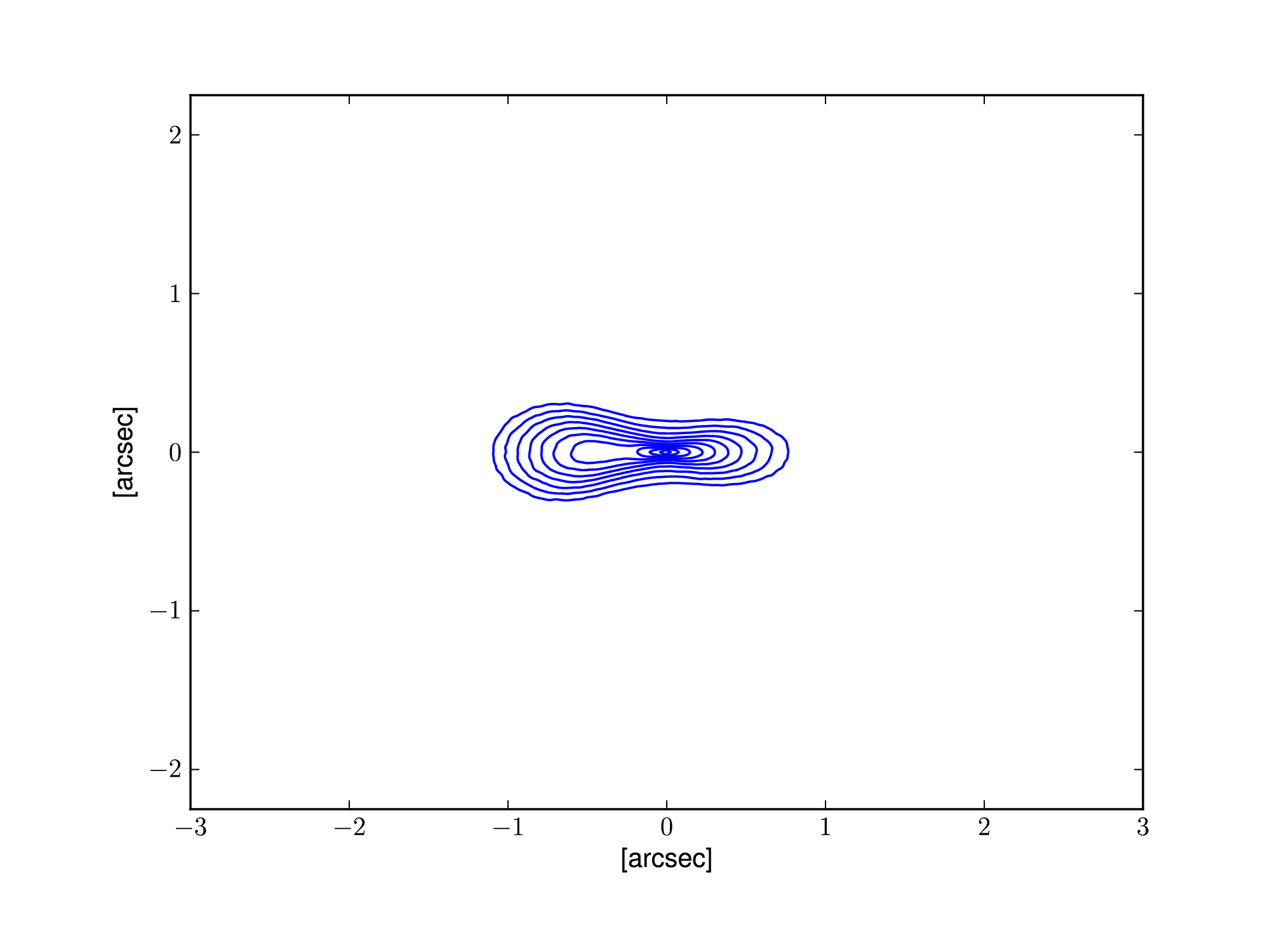}
  \caption{Face-on ($xy$, left) and edge-on ($xz$, right) projected surface
    brightness distributions of our best-fit model.
    Contours are spaced at 0.2 magnitudes.}
  \label{fig:faceedge}
\end{figure*}
Figure~\ref{fig:faceedge} shows the face-on and edge-on projected
density distributions of our best-fit model.  Our machinery fits a
density distribution which, when viewed face on, is broadly similar to
the density distribution adopted by \cite{salow04} in their
self-gravitating razor-thin models.  Apart from our neglect of self
gravity, the most significant difference between our models and theirs
is that ours have a secondary density peak around P2 and also have
significant thickness.

\begin{figure*}
  \includegraphics*[width=0.49\hsize, trim = 50 20 50 20]{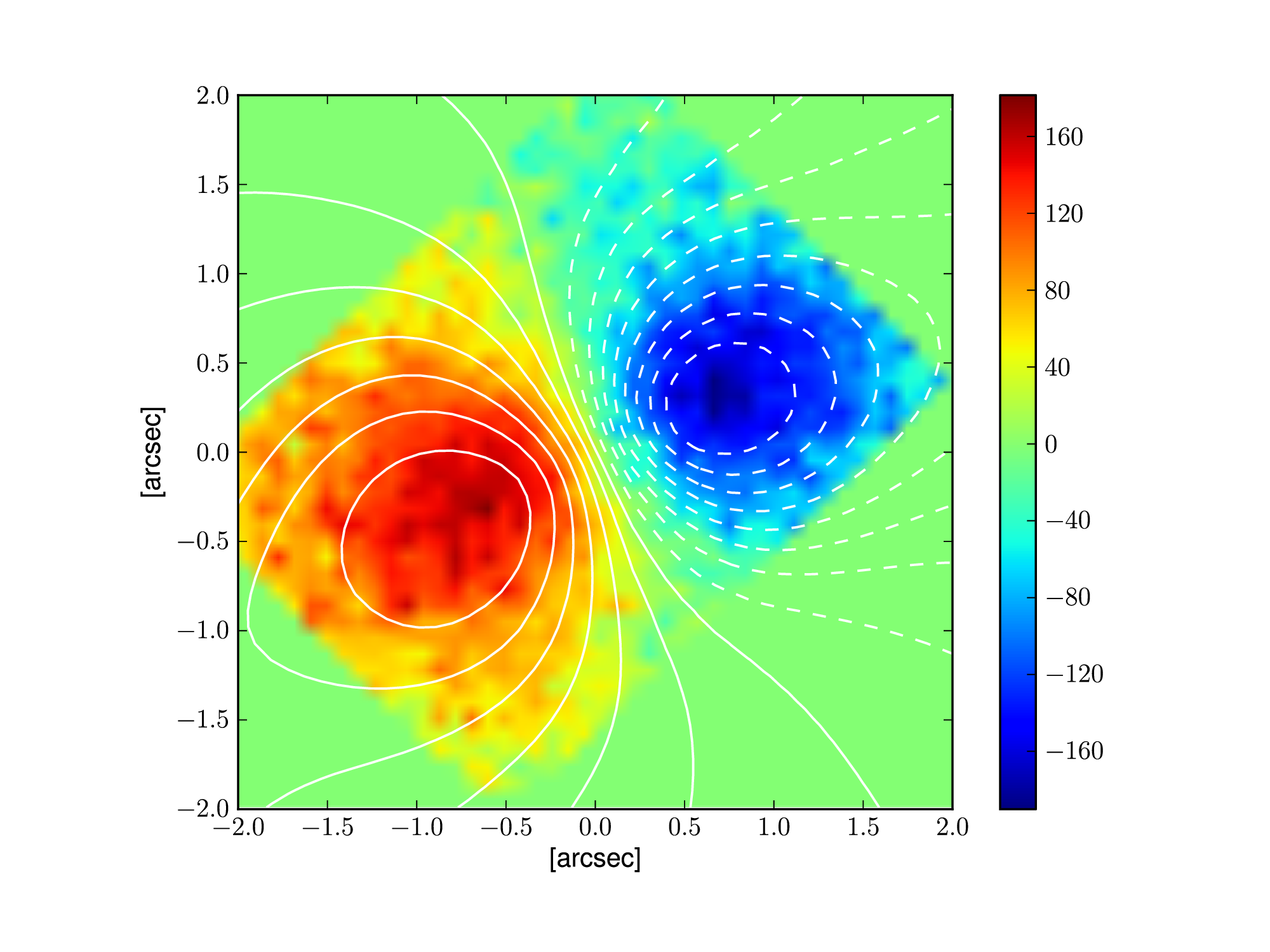}
  \includegraphics*[width=0.49\hsize, trim = 50 20 50 20]{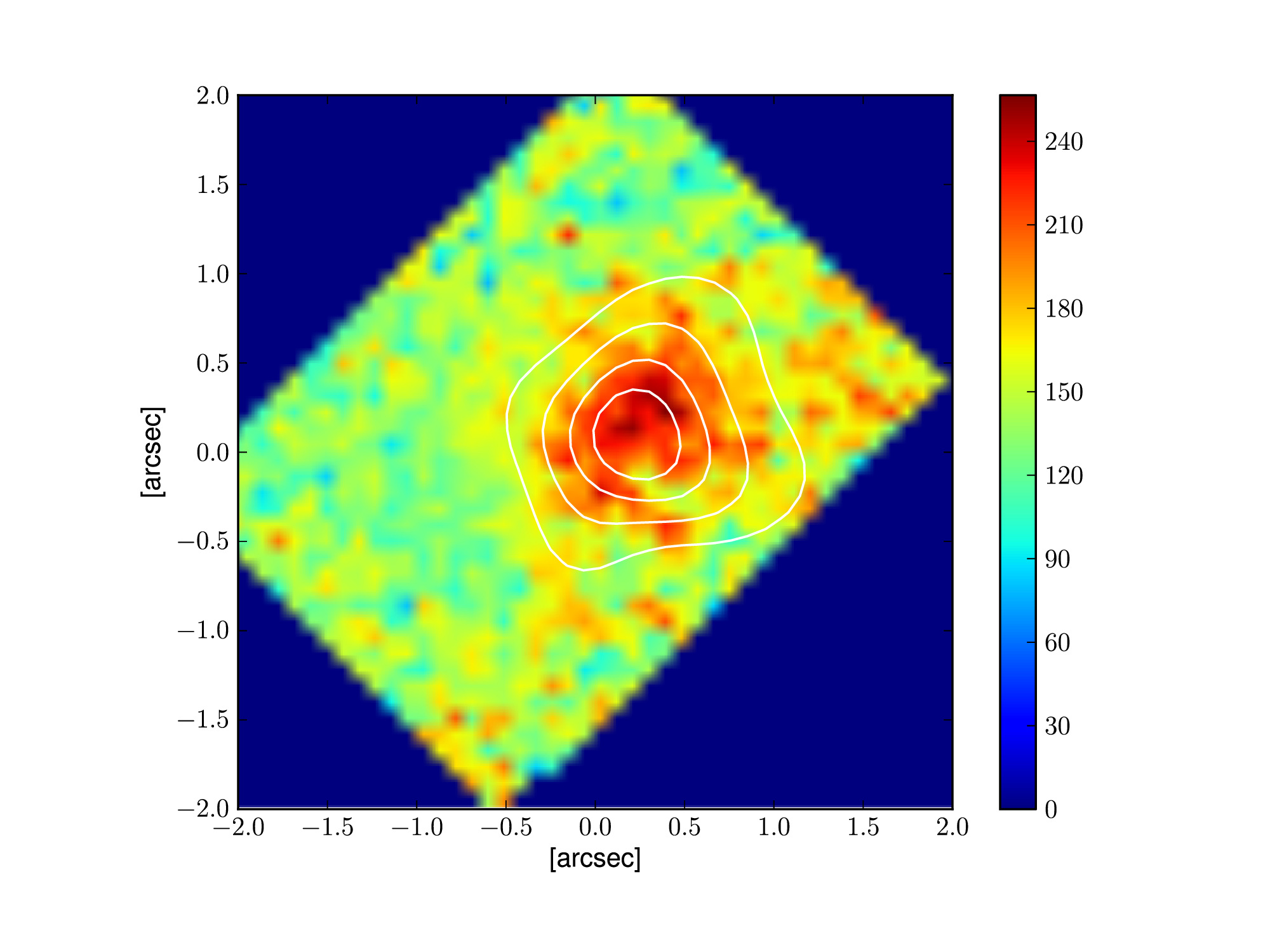}
  \caption{OASIS $V$ (left) and
    $\sigma$ (right) maps predicted by our models (white contours) compared to the observations from B01.  
    The contours in both panels are spaced $25\,\rm km\,s^{-1}$ apart.  The $V$ contours run from 0 to
    $\pm175\,\rm km\,s^{-1}$, $\sigma$ from 150 to $225\,\rm km\,s^{-1}$.}
  \label{fig:oasis}
\end{figure*}

Although the models we present in this section are not fit to the
OASIS maps, we can nevertheless compare our reconstructed models'
predictions against the real OASIS maps.  Figure~\ref{fig:oasis} shows
the results for the $M_\bullet=1.0\times10^8\,M_\odot$ model.  The
model reproduces the shape and orientation of velocity and dispersion
maps very well.  This provides an independent test of the
orientation angles and the broad-brush features of the DF inferred by our
models.

\begin{figure*}
  \centering
  \includegraphics*[width=0.49\hsize, trim = 20 30 0 80]{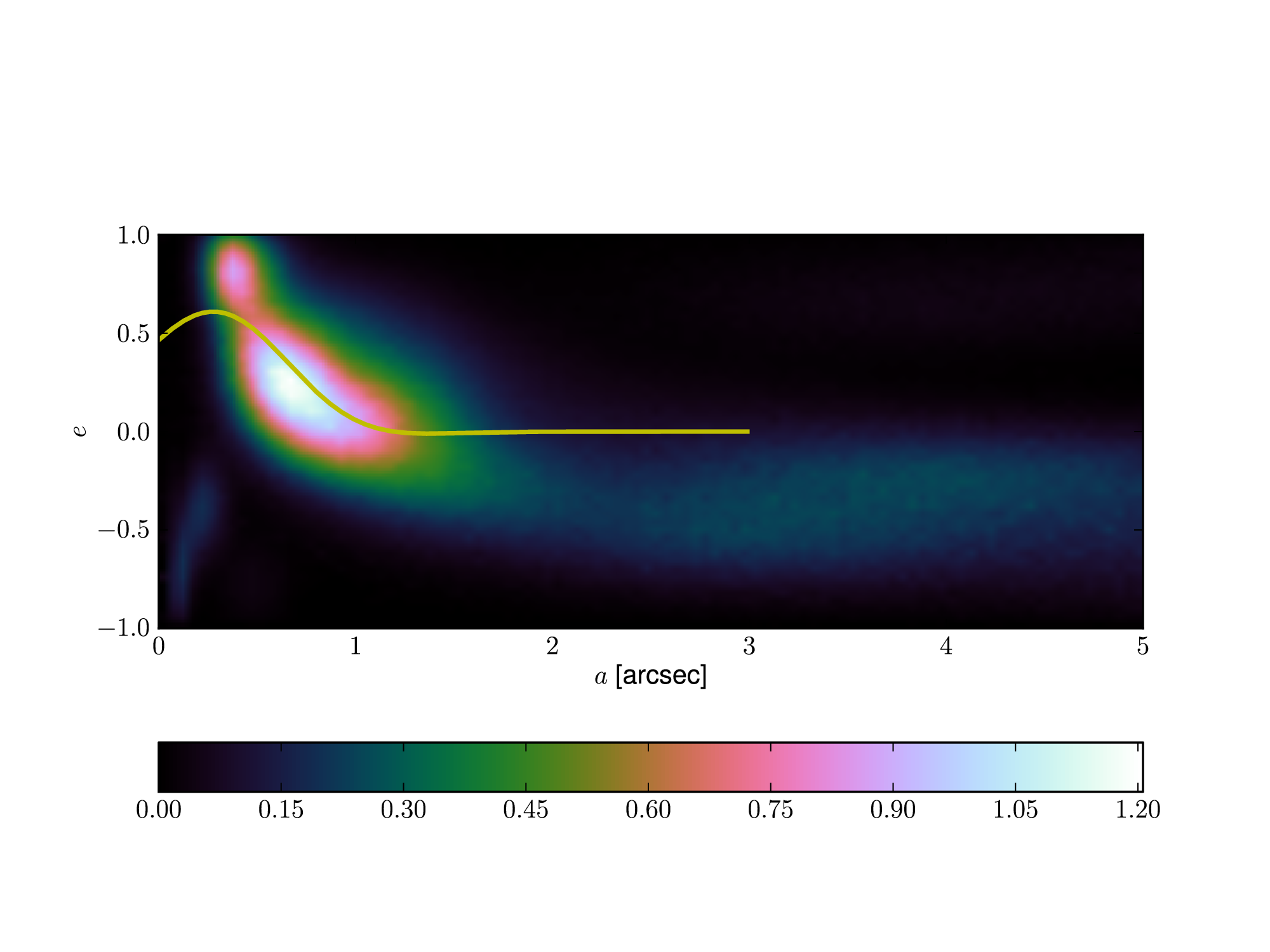}
  \includegraphics*[width=0.49\hsize, trim = 20 30 0 80]{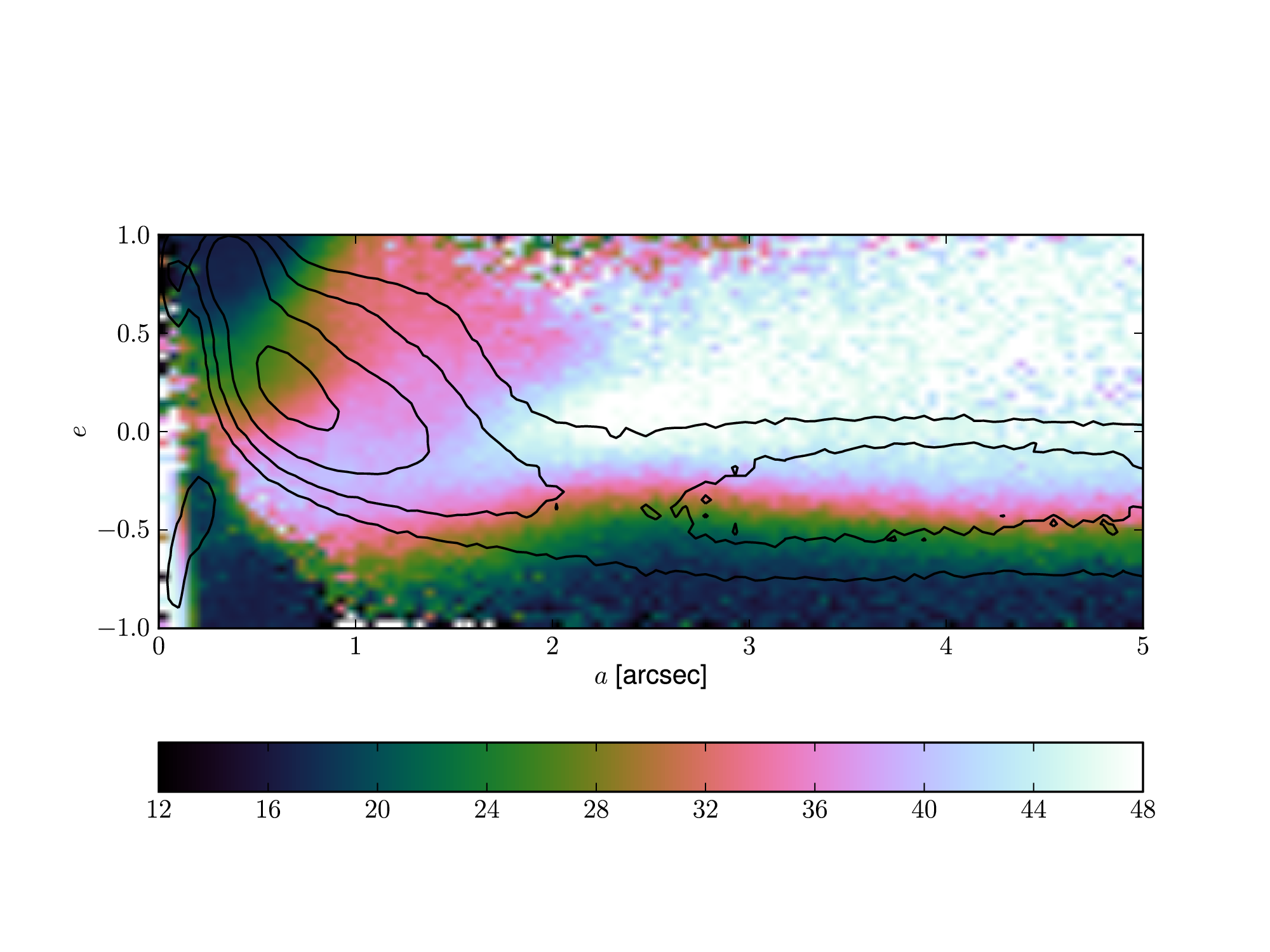}

  \includegraphics*[width=0.49\hsize, trim = 20 30 0 80]{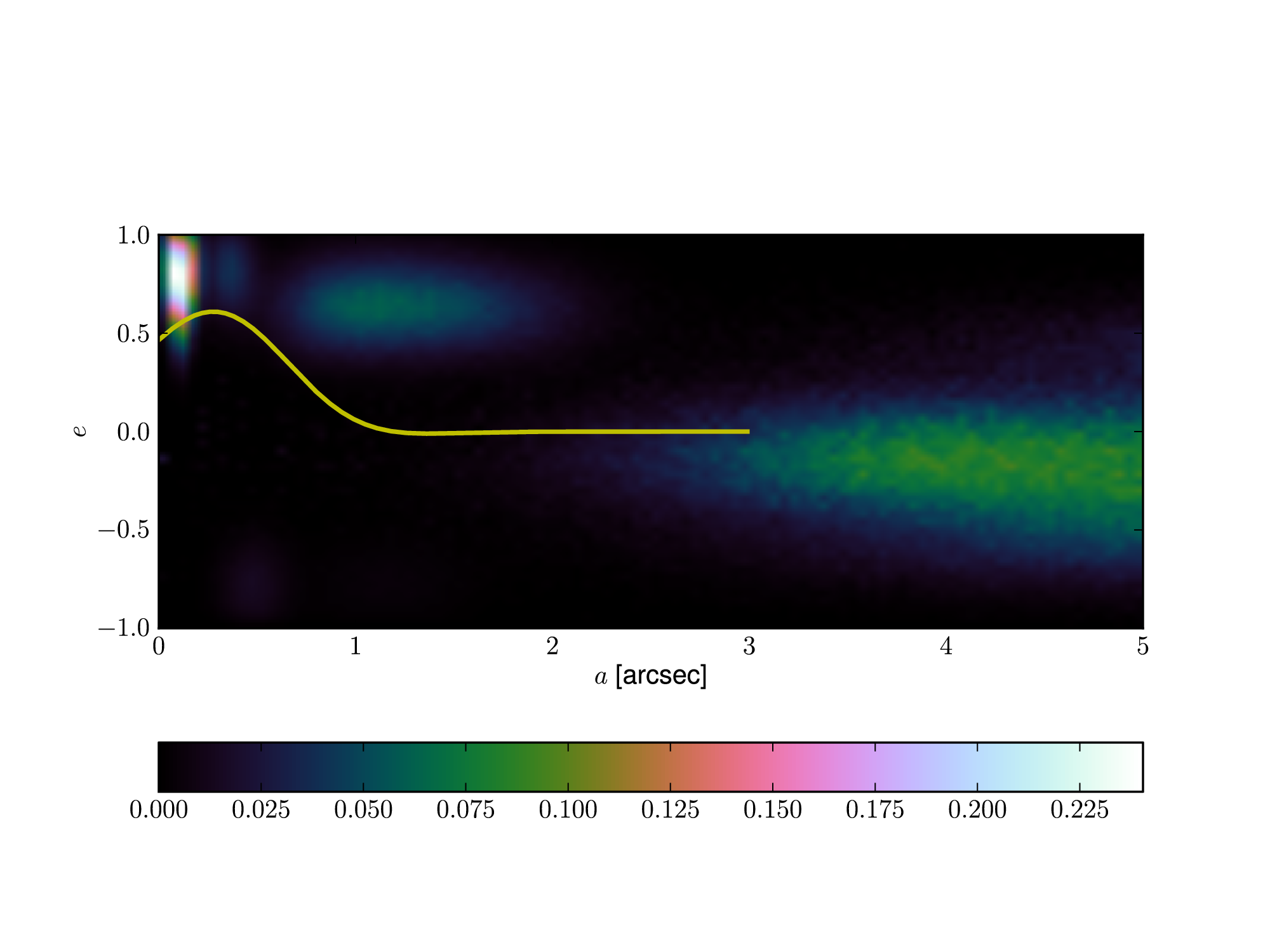}
  \includegraphics*[width=0.49\hsize, trim = 20 30 0 80]{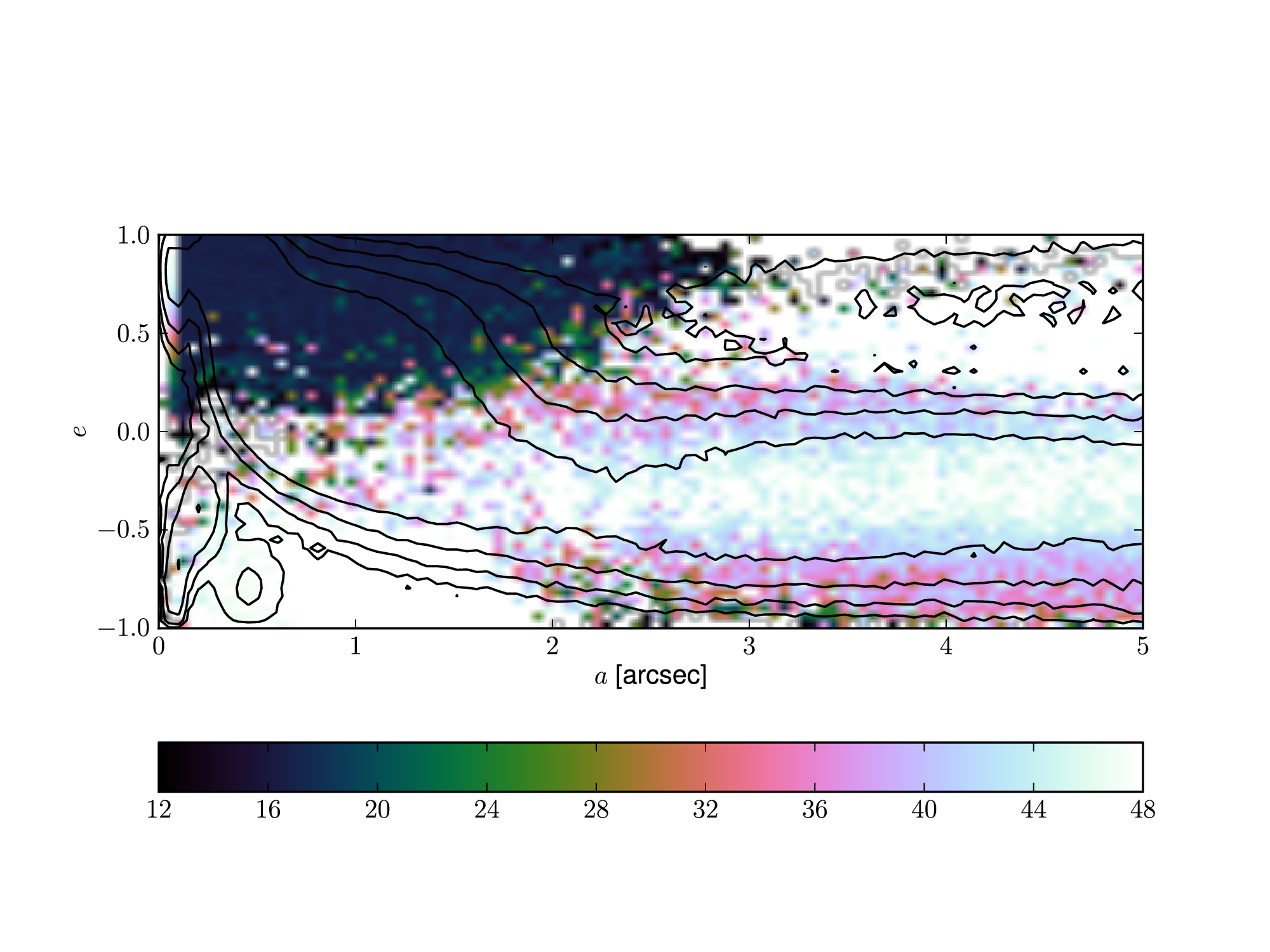}

  \caption{The distribution of orbits in a model that includes both
    prograde and retrograde orbits.  The top two panels show the DF
    \updated{$f(a,e)$}{$f(a,e_x)$} (left) and the rms inclination
    (right) for the prograde stars.  The bottom two panels show the
    corresponding distributions for the retrograde population.
  }
    \label{fig:DFplusminus}
\end{figure*}
\subsubsection{A brief experiment: retrograde orbits}

Finally, we note that one proposed origin of the eccentric disc in M31
is from an $m=1$ instability in an initially circular stellar disc
that contains a small counter-rotating population of stars
 \citep{touma02,jog09,kazandjian12}.  $N$-body simulations of this
instability \citep{kazandjian12} show that the dominant, prograde population
becomes the eccentric disc, while the minor, \updated{}{retrograde} population is
puffed up into a strongly triaxial distribution.  As a quick
experiment to test whether this is easily detectable, we have tried
doubling up our ring distribution by including in our models the retrograde ring
corresponding to each of the 720 prograde rings considered
\updated{}{above, giving a total of 1440 rings}.
This produces a noticeably better fit to the observations, with
$\chi^2=784$ instead of 812, \updated{}{with the fitting
  procedure picking out} three main counter-rotating
components (figure~\ref{fig:DFplusminus}) at $(a,e)=(0\farcs2,0.8)$,
$(1\farcs2,0.6)$ and $(3,-0.1)$.  None of these is easily identifiable
with the more diffuse component predicted by \cite{kazandjian12}, although
we suspect that that might be hard to detect with naive models such as
ours.  We interpret the innermost retrograde ring as suggesting that our
ring-based decomposition fails at radii $r<0\farcs2$, perhaps by not
being fine enough.  Similarly, we take the presence of the other two
rings as a hint that our bulge model could perhaps be improved.

\section{Conclusions}

We have constructed an eccentric disc models of the nucleus of M31 by
modelling it as a linear combination of fattened rings of stars
orbiting in a purely Keplerian potential.  Our models are an obvious
generalisation of those of PT03 and -- as expected -- are noticeably
(albeit perhaps not significantly) more successful at reproducing the
features observed in the inner arcsec or so of M31.

One of the most fundamental parameters of any model of this system is
its orientation.  Like PT03, we assume that the disc has biaxial
symmetry, but we find some differences in the Euler angles
$(\theta_a,\theta_i,\theta_l)$ that specify the orientation of these
symmetry axes.  Our values of $\theta_a = -34^\circ$ and $\theta_i =
57^\circ$ agree reasonably well with their $\theta_a=-34\fdg5$ and
$\theta_i = 54\fdg51$, but our value of $\theta_l=-35^\circ$ (which
directly controls the position angle on the sky) differs significantly
from their $\theta_l = -42\fdg8$, more than can be accounted for by
the effects of shot noise in either the models or in the distribution
of stars in the real disc.

Our orientation is also consistent with the N-body model of B01
($\theta_i=55^\circ \pm 5^\circ$ and $\theta_l=-36^\circ$) and the
models of \cite{salow04} ($\theta_i=63\fdg51 \pm 10\fdg80$ and assumed
$\theta_l=-33\fdg6$) and \cite{sambhus02} ($\theta_i=51\fdg54$ and
$\theta_l=27\fdg34$), though it should be noted all these models are
2d which imposes constraints on their geometry; \cite{sambhus02}
obtained their orientation by de-projecting the photometry of the disc
such that the outer isophotes became circular.  We note that these
values of $\theta_i$ found for the old, red distribution of stars are
close to the inclination of the young population in P3
($\theta_i=55^\circ \pm2^\circ$) measured by B+05.

The most interesting result of our models is the DF they infer from
the data, bearing in mind that they contain no prior ``wisdom'' about
which DFs are dynamically plausible.  The models
\updated{}{suggest} the presence of a
distinct, compact disc of red stars within 0\farcs15 of the black
hole.  This disc is anti-aligned (with eccentricity $e<0$) with
respect to the larger-scale eccentric disc (which has $e>0$).  Outside
this compact region the eccentricity distribution of our models is
very similar to PT03's.  The main difference between our models and
theirs is in the inclination distribution: our models fit an RMS
inclination profile $\sigma_I(a)$ that closely tracks the dispersion
in eccentricity, with $\sigma_I(a)/\sigma_e(a)\simeq2$.  This is
interesting: models of the collisional evolution of (circular) discs
predict that this ratio should be $\sim0.5$.  

Another difference between our models and PT03 is that we find
evidence for an anti-aligned feature $(e<0)$ at $a > 1\farcs5$.
However, the details of this feature depend on our assumed bulge
model, which merits further investigation.

Our models prefer black hole masses $M_\bullet$ of the order of
$1.0\times10^8\,M_\odot$; masses higher than about
$1.2\times10^8\,M_\odot$ are weakly ruled out.  Although it would be
possible to use our machinery to carry out a full scan of black hole
masses and orientation angles, followed by a systematic investigation
of the degeneracies in the DF, we believe that a more pressing task is
to include the self gravity of the disc.  The stars contribute a
significant fraction ($\sim20\%$) of the mass of the BH+eccentric disc
system, which means that it is dangerous to read too much into our
present, purely Keplerian models.  Past 2d
\citep[e.g.,]{bacon01,sambhus02,salow04} models have shown it is
possible to get plausibly good fits to the kinematics for large disc
masses. The space of 3d disc distributions that project to yield the
observed surface brightness profile is degenerate, but our orbital
ring system serves as a good starting point for this investigation.
\updated{}{Work on self-gravitating
self-consistent 3d disc models is now underway, and we hope that such
models will provide further insight into the origin of the eccentric
disc in M31 and possibly elsewhere} \citep{lauer05}.

\section*{Acknowledgments}
We thank the referee, S. Sridhar, for his careful reading of the
original version of this paper.

\label{lastpage}

\begin{thebibliography}{99}
\bibitem[\protect\citeauthoryear{Ajhar et al.}{1997}]{ajhar97} Ajhar, E. A., Lauer, T. R., Tonry, J. L., Blakeslee, J. P., Dressler, A., Holtzman, J. A., Postman, M.\ 1997, \aj, 114, 626 
\bibitem[\protect\citeauthoryear{Bacon et al.}{1994}]{bacon94} Bacon, R., Emsellem, E., Monnet, G., \& Nieto, J.~L.\ 1994, \aap, 281, 691 
\bibitem[\protect\citeauthoryear{Bacon et al.}{2001}]{bacon01} Bacon, R.; Emsellem, E.; Combes, F.; Copin, Y.; Monnet, G.; Martin, P.\ 2001, \aap, 371, 409 
\bibitem[\protect\citeauthoryear{Bender et al.}{2005}]{bender05} Bender, R., Kormendy, J., Bower, G., et al.\ 2005, \apj, 631, 280
\bibitem[\protect\citeauthoryear{Brown et al.}{1998}]{brown98} Brown, T.~M., Ferguson, H.~C., Stanford, S.~A., \& Deharveng, J.-M.\ 1998, \apj, 504, 113 
\bibitem[\protect\citeauthoryear{Cretton et al.}{1999}]{cretton99} Cretton, N., de Zeeuw, P.~T., van der Marel, R.~P., \& Rix, H.-W.\ 1999, \apjs, 124, 383 
\bibitem[\protect\citeauthoryear{Dressler \& Richstone}{1988}]{dressler88} Dressler, A., \& Richstone, D.~O.\ 1988, \apj, 324, 701
\bibitem[\protect\citeauthoryear{Jog \& Combes}{2009}]{jog09} Jog, C.~J., \& Combes, F.\ 2009, \physrep, 471, 75 
\bibitem[\protect\citeauthoryear{Kazandjian \& Touma}{2012}]{kazandjian12} Kazandjian, M.~V., \& Touma, J.~R.\ 2012, arXiv:1207.1108 
\bibitem[\protect\citeauthoryear{King et al.}{1995}]{king95} King, I.~R., Stanford, S.~A., \& Crane, P.\ 1995, \aj, 109, 164 
\bibitem[\protect\citeauthoryear{Kormendy}{1988}]{kormendy88} Kormendy, J.\ 1988, \apj, 325, 128 
\bibitem[\protect\citeauthoryear{Kormendy \& Bender}{1999}]{kormendy99} Kormendy, J., \& Bender, R.\ 1999, \apj, 522, 772 
\bibitem[\protect\citeauthoryear{Lauer et al.}{1993}]{lauer93} Lauer, T.~R., Faber, S.~M., Groth, E.~J., et al.\ 1993, \apj, 106, 1436 
\bibitem[\protect\citeauthoryear{Lauer et al.}{1998}]{lauer98} Lauer, T.~R., Faber, S.~M., Ajhar, E.~A., Grillmair, C.~J., \& Scowen, P.~A.\ 1998, \apj, 116, 2263 
\bibitem[\protect\citeauthoryear{Lauer et al.}{2005}]{lauer05} Lauer,  T.~R., et al., 2005, \aj 129 2138
\bibitem[\protect\citeauthoryear{Lauer et al.}{2012}]{lauer12} Lauer, T.~R., Bender, R., Kormendy, J., Rosenfield, P., \& Green, R.~F.\ 2012, \apj, 745, 121 
\bibitem[\protect\citeauthoryear{Lawson \& Hanson}{1974}]{lawson74} Lawson, C.~L., \& Hanson, R.~J.\ 1974, Prentice-Hall Series in Automatic Computation, Englewood Cliffs: Prentice-Hall, 1974, 
\bibitem[\protect\citeauthoryear{Light et al.}{1974}]{light74} Light, E.~S., Danielson, R.~E., \& Schwarzschild, M.\ 1974, \apj, 194, 257 
\bibitem[\protect\citeauthoryear{Peiris \& Tremaine}{2003}]{peiris03} Peiris, H.~V., \& Tremaine, S.\ 2003, \apj, 599, 237
\bibitem[\protect\citeauthoryear{Salow \& Statler}{2001}]{salow01} Salow, R.~M., \& Statler, T.~S.\ 2001, \apj, 551, L49  
\bibitem[\protect\citeauthoryear{Salow \& Statler}{2004}]{salow04} Salow, R.~M., \& Statler, T.~S.\ 2004, \apj, 611, 245 
\bibitem[\protect\citeauthoryear{Sambhus \& Sridhar}{2002}]{sambhus02} Sambhus, N., \& Sridhar, S.\ 2002, \aap, 388, 766 
\bibitem[\protect\citeauthoryear{Statler}{1999}]{statler99} Statler, T.~S.\ 1999, \apj, 524, L87 
\bibitem[\protect\citeauthoryear{Statler et al.}{1999}]{statleretal99} Statler, T.~S., King, I.~R., Crane, P., \& Jedrzejewski, R.~I.\ 1999, \aj, 117, 894 
\bibitem[\protect\citeauthoryear{Stewart \& Ida}{2000}]{stewart00} Stewart G.~R., Ida  S., 2000, Icarus, 143, 28
\bibitem[\protect\citeauthoryear{Touma}{2002}]{touma02} Touma, J.~R.\ 2002, \mnras, 333, 583
\bibitem[\protect\citeauthoryear{Tremaine}{1995}]{tremaine95} Tremaine, S.\ 1995, \apj, 110, 628
\bibitem[\protect\citeauthoryear{van der Marel \& Franx}{1993}]{vdm93} van der Marel, R.~P., \& Franx, M.\ 1993, \apj, 407, 525
\bibitem[\protect\citeauthoryear{van der Marel et al.}{1994}]{vdm94} van der Marel, R. P.; Rix, H. W.; Carter, D.; Franx, M.; White, S. D. M.; de Zeeuw, T.\ 1994, \mnras, 268, 521 

\end{thebibliography}
\end{document}